\documentclass[reprint,nofootinbib,aps]{revtex4-1}
\usepackage{amsmath,amssymb,amsbsy,amstext,amsthm,simplewick,amsfonts,slashed} 
\usepackage{physics}
\usepackage{graphicx} 
\usepackage{dcolumn} 
\usepackage{bm} 
\usepackage{comment}
\usepackage{enumitem}
\usepackage{hyperref}
\usepackage{multirow}
\hypersetup{
	colorlinks = true,
	linkcolor = blue,
	citecolor = blue,
	urlcolor = blue
}

\def\beq{\begin{equation}}
\def\eeq{\end{equation}}

\newcommand{\veck}{\mathbf{k}}

\newcommand{\mphieff}{m_\text{$\phi$,eff}}
\newcommand{\mchieff}{m_\text{$\chi$,eff}}
\newcommand{\xcfo}{x_\text{cfo}}
\newcommand{\xcfotilde}{\tilde{x}_\text{cfo}}

\usepackage[normalem]{ulem}

\begin{document}
\title{WIMP Meets ALP: Coherent Freeze-Out of Dark Matter}
\author{Steven Ferrante}
\email{sef87@cornell.edu}
\author{Maxim Perelstein}
\email{m.perelstein@cornell.edu}
\author{Bingrong Yu}
\email{bingrong.yu@cornell.edu}
\affiliation{Department of Physics, LEPP, Cornell University, Ithaca, NY 14853, USA}

\begin{abstract}

We consider the cosmological history of a weakly interacting massive particle (WIMP) coupled to a light axion-like particle (ALP) via a quadratic coupling. Although the coupling is too feeble to thermalize the ALP, coherent forward scattering between the two sectors induces temperature-dependent mass shifts that substantially modify both WIMP freeze-out and ALP misalignment dynamics, giving rise to a novel \emph{coherent freeze-out} mechanism.
At high temperatures, the WIMP thermal bath spontaneously breaks the symmetry of the ALP potential, displacing the field to a new vacuum. The resulting back-reaction reduces the WIMP effective mass and significantly delays its freeze-out. Depending on the strength of the coupling, symmetry restoration occurs via either a first-order phase transition (FOPT) or a crossover. In the FOPT regime, dark matter consists solely of WIMPs, whose delayed freeze-out permits annihilation cross sections up to three orders of magnitude above the standard value, while still yielding the correct relic density. In the crossover regime, both WIMP and ALP can contribute to dark matter. Remarkably, we find an ``ALP miracle": a Planck-suppressed quadratic coupling yields an ALP abundance comparable to the observed dark matter density, largely independent of its initial displacement and mass.
\end{abstract}

\maketitle

\section{Introduction}
The particle nature of dark matter (DM) remains unknown. Two well-motivated candidates beyond the Standard Model (SM) are weakly interacting massive particles (WIMPs) and axion-like particles (ALPs) (see~\cite{Cirelli:2024ssz,Marsh:2015xka} for recent reviews).
They inhabit very different mass ranges and arise from distinct theoretical frameworks, and their cosmological production mechanisms likewise differ sharply.
WIMPs begin in thermal equilibrium and obtain their relic abundance through freeze-out, a process largely insensitive to ultraviolet (UV) physics.
ALPs, by contrast, never thermalize due to their extremely weak couplings; their relic abundance is instead determined by the initial field displacement at the onset of coherent oscillation, typically set by inflationary fluctuations or other high-scale dynamics.

In this work, we raise a generic question: if both a WIMP and an ALP are present in the fundamental Lagrangian, what are the cosmological consequences of an interaction between them?
Since the coupling is typically too weak to thermalize the ALP, one might expect their dynamics to remain independent, as no significant momentum exchange occurs. However, this intuition fails. The ALP behaves as a coherent, homogeneous field, while the WIMP population forms an approximately uniform thermal bath. Coherent forward scattering between the two backgrounds induces mean-field corrections to the dispersion relations of both species: the ALP potential is deformed by the thermal WIMP bath, and the WIMP effective mass is shifted by the ALP background. These medium effects lead to coupled, nontrivial dynamics in their cosmological evolution. The goal of this paper is to show that even an extremely weak (Planck-suppressed) coupling between the two sectors leads to striking cosmological consequences.

To illustrate the mechanism, we consider a simple model in which WIMP and ALP fields interact through an effective quadratic coupling (see Eq.~(\ref{eq:Leff})). The thermal bath of WIMPs induces a temperature-dependent correction to the ALP potential, spontaneously breaking its discrete symmetry at high temperatures and restoring it at low temperatures. Such an ``inverse phase transition''
occurs not only in many particle-physics models~\cite{Mohapatra:1979vr,Dvali:1995cc,Pietroni:1996zj,Riotto:1997tf,Espinosa:2004pn,Shakya:2025mdh}, but also in condensed-matter systems~\cite{PhysRevE.72.046107}.
In the broken phase, the ALP field is displaced to a new vacuum, whose back-reaction reduces the effective WIMP mass and therefore affects its freeze-out. At the same time, ALP coherent oscillations only begin after the phase transition, with the initial value of the field fixed by the high-temperature dynamics, in contrast with the usual misalignment mechanism.   

This setup leads to several key consequences:
(i) the symmetry restoration of the ALP potential is governed by a single dimensionless parameter, which determines whether the system undergoes a first-order phase transition (FOPT) or a smooth crossover;
(ii) in the FOPT regime, the WIMP may remain in thermal equilibrium far longer than in the standard picture, leading to a substantial enhancement of its relic abundance -- the scenario we call ``coherent freeze-out";
(iii) in the crossover regime, the ALP obtains a relic abundance that deviates from the usual misalignment prediction and, remarkably, is insensitive to both the initial ALP field value and the ALP mass.

Modifications to the dynamics of a light scalar in the presence of a thermal bath have been discussed in several contexts (see, e.g.,~\cite{Piazza:2010ye,Moroi:2013tea,Lillard:2018zts,Ramazanov:2021eya,Batell:2021ofv,Cyncynates:2024bxw, Cyncynates:2024ufu,Banerjee:2025nvs,Graham:2025gtd}).
The crucial distinction in our analysis is that we fully incorporate the back-reaction of the scalar on the thermal bath -- an effect that plays a central role in our mechanism, especially in the FOPT regime.

\section{Dynamics}

\begin{figure*}[t!]
\centering
\includegraphics[width=0.53\textwidth]
{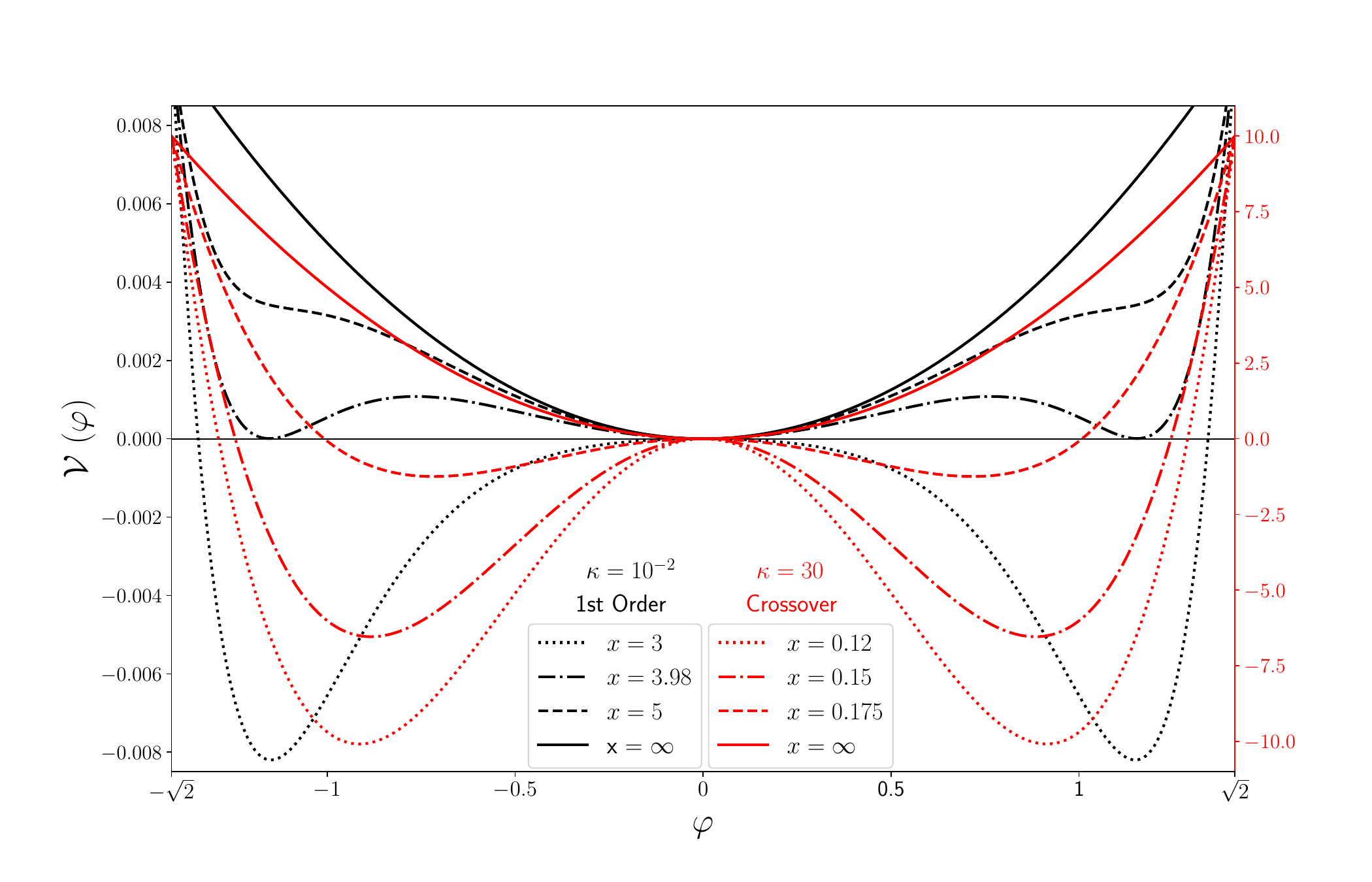}
\includegraphics[width=0.46\textwidth]
{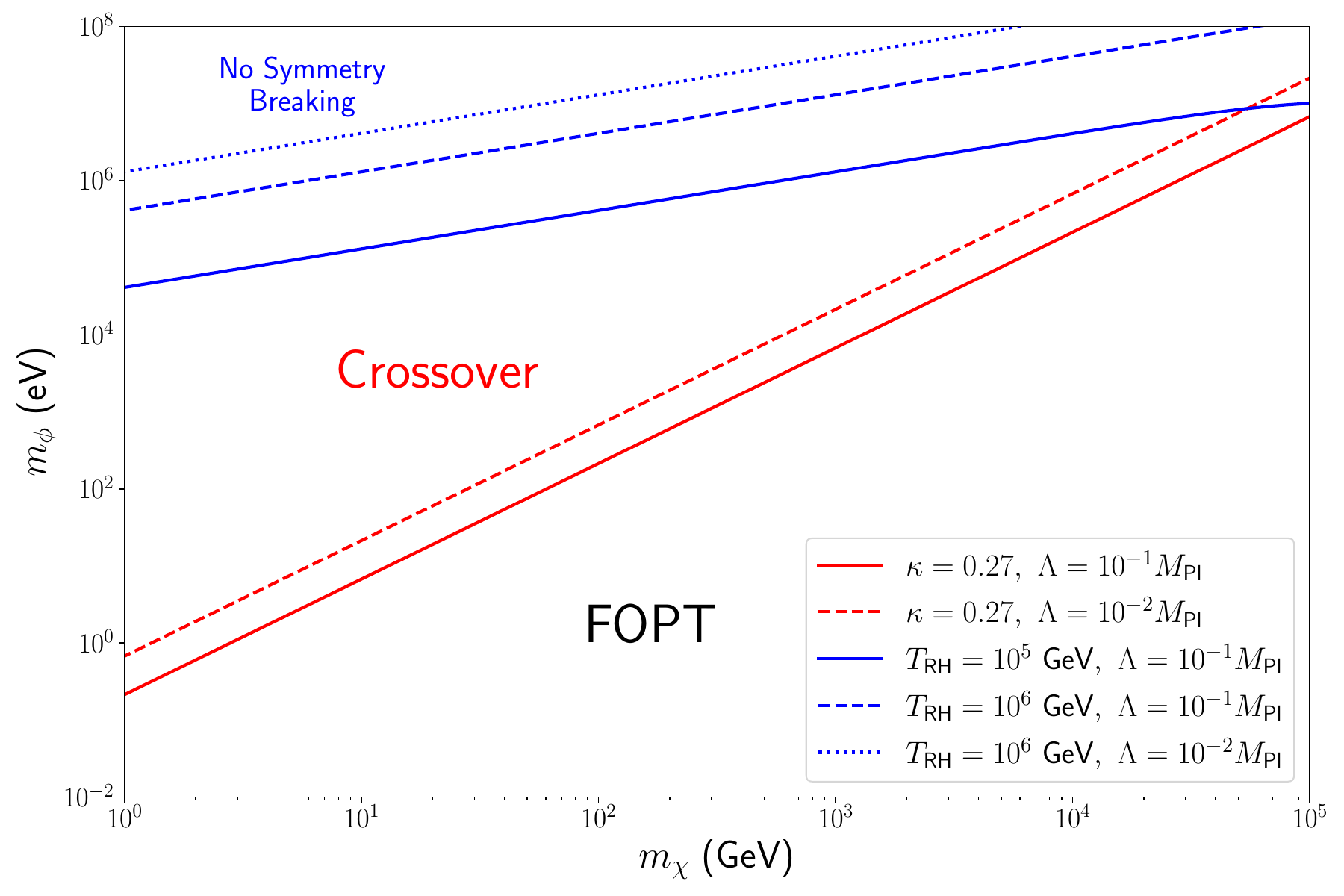}
\caption{\label{fig:potential}
\textbf{Left:} The normalized ALP potential~(\ref{eq:potential}) in a thermal bath of WIMPs at several temperatures.  
For $\kappa < \kappa_c \approx 0.27$ the phase transition is first order (black curves), while for $\kappa > \kappa_c$ it becomes a crossover (red curves).  
\textbf{Right:} Phase diagram in the $(m_{\chi}, m_{\phi})$ plane.  
Below the $\kappa = \kappa_c$ contour (red lines), symmetry restoration proceeds through a first-order phase transition; above it, the transition is a crossover.  
Symmetry restoration further requires that the symmetry be broken at the onset of radiation domination, $T_{\rm RH}$.  
The condition $\kappa = K_{1}(x_{\rm RH})/x_{\rm RH}$ (blue curves), with $x_{\rm RH} \equiv m_\chi/T_{\rm RH}$, marks the boundary above which the symmetry is never broken and no transition occurs.}
\end{figure*}

We begin with the following effective coupling,
\begin{align}
{\mathcal L}_{\rm eff} \supset\frac{1}{\Lambda} \bar{\chi}\chi\frac{\phi^2}{2}\;,
\label{eq:Leff}
\end{align}
where $\Lambda$ is the cutoff scale, $\chi$ is a fermionic WIMP (either Dirac or Majorana), and $\phi$ is a light real scalar. For convenience, we refer to $\phi$ as the ALP hereafter, although it may be CP-even or CP-odd. The model contains three hierarchical scales, $\Lambda \gg m_\chi \gg m_\phi$, with $m_\chi$ and $m_\phi$ denoting the physical zero-temperature masses of $\chi$ and $\phi$, respectively. For simplicity, we do not include higher-order terms in $\phi$ in the zero-temperature potential.
The coupling in (\ref{eq:Leff}) is generic and can arise in a wide range of UV completions. We take a definite sign for the interaction, which is essential for our analysis. Scalar-fermion interactions with the same structure as (\ref{eq:Leff}), including the sign, appear in familiar contexts. For example, this is the structure of the leading non-derivative interaction between nucleons and pions in the SM, or between nucleons and the QCD axion in minimal extensions, both of which follow directly from chiral effective theory~\cite{Epelbaum:2008ga,Machleidt:2011zz,Hook:2017psm,Bigazzi:2019hav,DiLuzio:2020wdo,Fukuda:2021drn}.
Throughout, we assume the coupling is small enough that $\phi$ never thermalizes (neither chemically nor kinematically), which requires $\Lambda \gg \sqrt{T_{\rm RH} M_{\rm Pl}}$, where $T_{\rm RH}$ is the reheating temperature of the Universe and $M_{\rm Pl} \approx 1.2 \times 10^{19}~{\rm GeV}$ is the Planck mass.

Even without momentum exchange, coherent forward scattering modifies the propagation of both species. In the region of interest, $\chi$ is in thermal equilibrium with the SM.
The thermal $\chi$ background shifts the ALP mass, and the classical ALP field shifts the WIMP mass,
\begin{align}
\mphieff^2 &= m_\phi^2 - \frac{\langle \bar{\chi}\chi \rangle_T}{\Lambda}\;, 
\label{eq:mphieff}\\
\mchieff &= \left|m_\chi -\frac{\phi^2}{2\Lambda}\right|\,,\label{eq:mchieff}
\end{align}
where 
\begin{align}
\langle \bar{\chi}\chi \rangle_T = g_\chi\int \frac{{\rm d}^3\veck}{(2\pi)^3}\frac{\mchieff}{E_\veck}f_\chi(\veck)\;,\label{eq:chibarchi}
\end{align}
is the thermally averaged number density, 
with $E_\veck\equiv \sqrt{\veck^2+\mchieff^2}$ and $g_\chi = 4 (2)$ if $\chi$ is Dirac (Majorana). Crucially, the correction to $m_\phi$ depends on $\mchieff$, which itself depends on $\phi$; the dynamics of the two sectors are therefore \emph{coupled}.

For analytic control, we adopt a Maxwell-Boltzmann (MB) distribution, 
$f_\chi = e^{-E_\veck/T}$. 
(A more complete treatment -- including quantum-statistical effects and 
thermal corrections beyond quadratic order -- can be implemented numerically; see 
Appendix~\ref{app:fullthermal}. These refinements do not modify the qualitative 
results.)
This leads to the ALP potential

\begin{align}
V(\phi, T)  = 
    \frac{g_{\chi}}{2\pi^{2}}
    m_{\chi}^{4}
 \, \mathcal{V}(\varphi, x)\;,    
\end{align}
with the normalized potential
\begin{align}
\mathcal{V}(\varphi, x) &= 
    -\frac{\varphi^{2}}{2}
   \frac{\gamma^{2} 
    K_{1}
    (\gamma x)}{x}
    + 
    \kappa\frac{\varphi^{2}}{2}\;,\label{eq:potential}  
\end{align}
where $K_1$ is the modified Bessel function. Here we have defined $x\equiv m_\chi/T$, the normalized ALP field $\varphi \equiv \phi/\sqrt{m_{\chi}\Lambda}$, the boost factor $\gamma \equiv |1-\varphi^2/2|$, and
a dimensionless parameter
\begin{align}
\kappa \equiv \frac{2\pi^2}{g_{\chi}}\frac{m_{\phi}^{2}\Lambda}{m_{\chi}^{3}}\;.\label{eq:kappadef} 
\end{align}

The first term in (\ref{eq:potential}) encodes the thermal effect of the WIMP bath, while the second is the bare ALP mass. The resulting potential, shown in the left panel of Fig.~\ref{fig:potential}, contains a term quartic in $\varphi$ due to the back-reaction on $\mchieff$, ensuring it is bounded from below.
At early times ($\gamma x\ll1$), the thermal term acts like a negative $T^{2}$ mass and spontaneously breaks the $Z_2$ symmetry ($\phi \to -\phi$); at late times, the bare mass dominates and restores it. The order of this transition is set by a critical coupling $\kappa_{c} \approx 0.27$, which can be calculated in Ginzburg-Landau theory (see Appendix \ref{app:orderPT}) For $\kappa<\kappa_{c}$, the transition is first-order, and for $\kappa>\kappa_{c}$ it is a crossover.
We truncate the potential at $|\varphi|=\sqrt{2}$, where $\mchieff\to 0$ and higher-order terms become important (see Appendix~\ref{app:fullthermal}). Since the field starts with negligible velocity after inflation, $|\varphi|$ never exceeds this bound.

In the radiation era, the ALP field obeys the following equation of motion (EOM): 
\begin{widetext}
\begin{align}
&\varphi'' + \frac{2}{x}\varphi' - \eta x \left(1-\frac{\varphi^2}{2}\right)\left[\left(1-\varphi^2\right) K_1(\gamma x)+ \frac{\varphi^2}{2}\gamma x K_0(\gamma x)
\right] \varphi+ \eta \kappa x^2 \varphi = 0\;,
\label{eq:EOMvarphi}
\end{align}
\end{widetext}
where prime denotes the derivative with respect to $x$ and 
\begin{align}
\eta \equiv \frac{g_\chi}{2\pi^2}\frac{1}{1.66^2 g_*}\frac{M_{\rm Pl}^2}{m_\chi \Lambda}\;,\label{eq:etadef}    
\end{align}
with $g_*$ the number of relativistic degrees of freedom. The EOM (\ref{eq:EOMvarphi}) contains only two free parameters ($\kappa$ and $\eta$), and its solutions are studied in detail in Appendix~\ref{app:solution}. Depending on $\kappa$, the resulting parameter space falls into three classes (see right panel of Fig.~\ref{fig:potential}):

\begin{enumerate}
    \item \textbf{No symmetry breaking:} $\kappa > K_1(x_{\rm RH})/x_{\rm RH}$ with $x_{\rm RH}\equiv m_\chi/T_{\rm RH}$. Symmetry is never broken, and the two sectors decouple.

    \item \textbf{Crossover:} $\kappa_c<\kappa<K_1(x_{\rm RH})/x_{\rm RH}$. Symmetry is broken at high $T$ and restored smoothly at low $T$. 
    
    \item \textbf{FOPT:}
    $\kappa<\kappa_c$ and $\kappa<K_1(x_{\rm RH})/x_{\rm RH}$. Symmetry is broken at high $T$ and restored at low $T$ through a first-order transition. 
\end{enumerate}
In what follows, we focus on the latter two scenarios, where the WIMP and ALP exhibit genuinely coupled dynamics. 

\section{Phenomenology}
Now we turn to the DM phenomenology in our mechanism and distinguish two regimes.
In the FOPT regime ($\kappa \lesssim 0.27$), the ALP behaves as a spectator field and does not contribute to the relic abundance. In the crossover regime ($\kappa \gtrsim 0.27$), the ALP provides a second DM component whose abundance is largely insensitive to its initial field displacement and mass.

\subsection{First-order phase transition regime} 

At high temperatures, the ALP potential develops a symmetry-breaking minimum $\varphi_* \neq 0$ (see Fig.~\ref{fig:potential}). We define $\xcfotilde$ as the time when this local minimum disappears:
\begin{align}
  \frac{\partial{\cal V}(\varphi,\xcfotilde)}{\partial \varphi} \Bigg|_{\varphi=\varphi_*}=\frac{\partial^2{\cal V}(\varphi,\xcfotilde)}{\partial \varphi^2} \Bigg|_{\varphi=\varphi_*} = 0\;,    
\end{align}
which determines both $\varphi_*$ and $\xcfotilde$ as functions of $\kappa$. In the limit of $\kappa \ll 1$ and $x\gg 1$, the symmetry-breaking vacuum evolves as (see Appendix~\ref{app:WIMPdynamics})
\begin{align}
\varphi_* = \sqrt{2\left(1-\frac{c_1}{x}\right)}\;,\label{eq:phistar}    
\end{align}
where $c_1\approx 1.33$ is the root of $c_1 K_0(c_1) = K_1(c_1)$, and $\xcfotilde$ follows the simple scaling
\begin{align}
\xcfotilde = c_2\,\kappa^{-1/3}\;, \quad c_2 \approx 0.83\;.
\label{eq:xcfotildekappa}
\end{align}
Here we use the full thermal potential presented in Appendix~\ref{app:fullthermal} to obtain  Eq.~(\ref{eq:xcfotildekappa}), providing a more accurate quantitative description.
For $x<\xcfotilde$, the tunneling rate is highly suppressed until very close to $\xcfotilde$, so the field remains trapped at $\varphi_*$. It returns to $\varphi=0$ only when $x> \max\left\{\xcfotilde,x_{\rm osc}\right\}$, where $x_{\rm osc}\sim m_\chi/\sqrt{m_\phi M_{\rm Pl}}$ marks the onset of ALP oscillations. 

We now show how the above dynamics affect WIMP freeze-out. The Boltzmann equation governing the WIMP number density is
\begin{align}
\dot{n}_\chi + 3 Hn_\chi = - \langle \sigma v\rangle_{\rm eff} \left(n_\chi^2-n_{\chi,{\rm eq}}^2\right). \label{eq:Boltzmann-maintext}
\end{align}
The ALP background reduces the WIMP effective mass, modifying both $\langle \sigma v\rangle_{\rm eff}$ and the equilibrium distribution:
\begin{align}
n_{\chi,{\rm eq}} = \frac{g_\chi}{2\pi^2} \mchieff^2 T K_2(\mchieff/T)\;.\label{eq:neq}    
\end{align}
When $\varphi =\varphi_*$, Eq.~(\ref{eq:phistar}) implies 
\begin{align}
\mchieff/T = (1-\varphi_*^2/2)\,x = c_1 \approx 1.33\;.
\end{align}
So the argument of the Bessel function remains order unity even at $x\gg 25$. This allows the WIMP to stay in equilibrium far longer than in the standard freeze-out scenario.

\begin{figure}[t!]
\centering
\includegraphics[width=3.6in]
{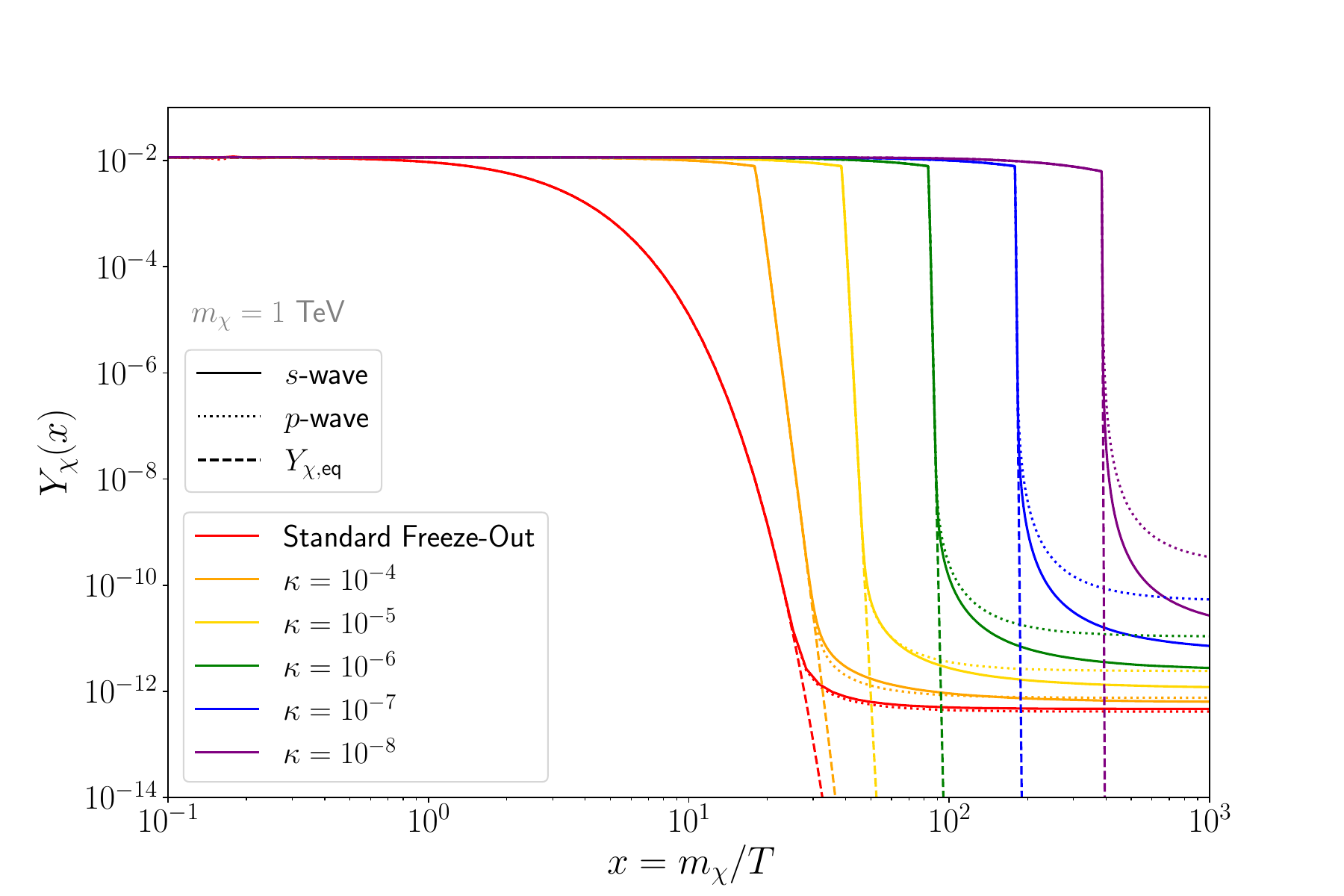}
\caption{\label{fig:abundance}
WIMP yield $Y_\chi \equiv n_\chi/s$ as a function of inverse temperature in the coherent freeze-out scenario. 
Solid (dotted) curves show the numerical solution of the Boltzmann equation~(\ref{eq:Boltzmann-maintext}) for $s$-wave ($p$-wave) annihilation, while dashed curves show the corresponding equilibrium yield~(\ref{eq:neq}). 
For comparison, the standard freeze-out result is shown in red. 
The benchmark parameters used are $g_\chi =2$, $m_\chi = 1~\mathrm{TeV}$, with $\sigma_0 = 2.2\times10^{-26}~\mathrm{cm^3/s}$ for $s$-wave and $\sigma_1 = 1.9\times10^{-25}~\mathrm{cm^3/s}$ for $p$-wave domination.
}

\end{figure}

\begin{figure*}[t!]
\centering
\includegraphics[width=0.8\textwidth]
{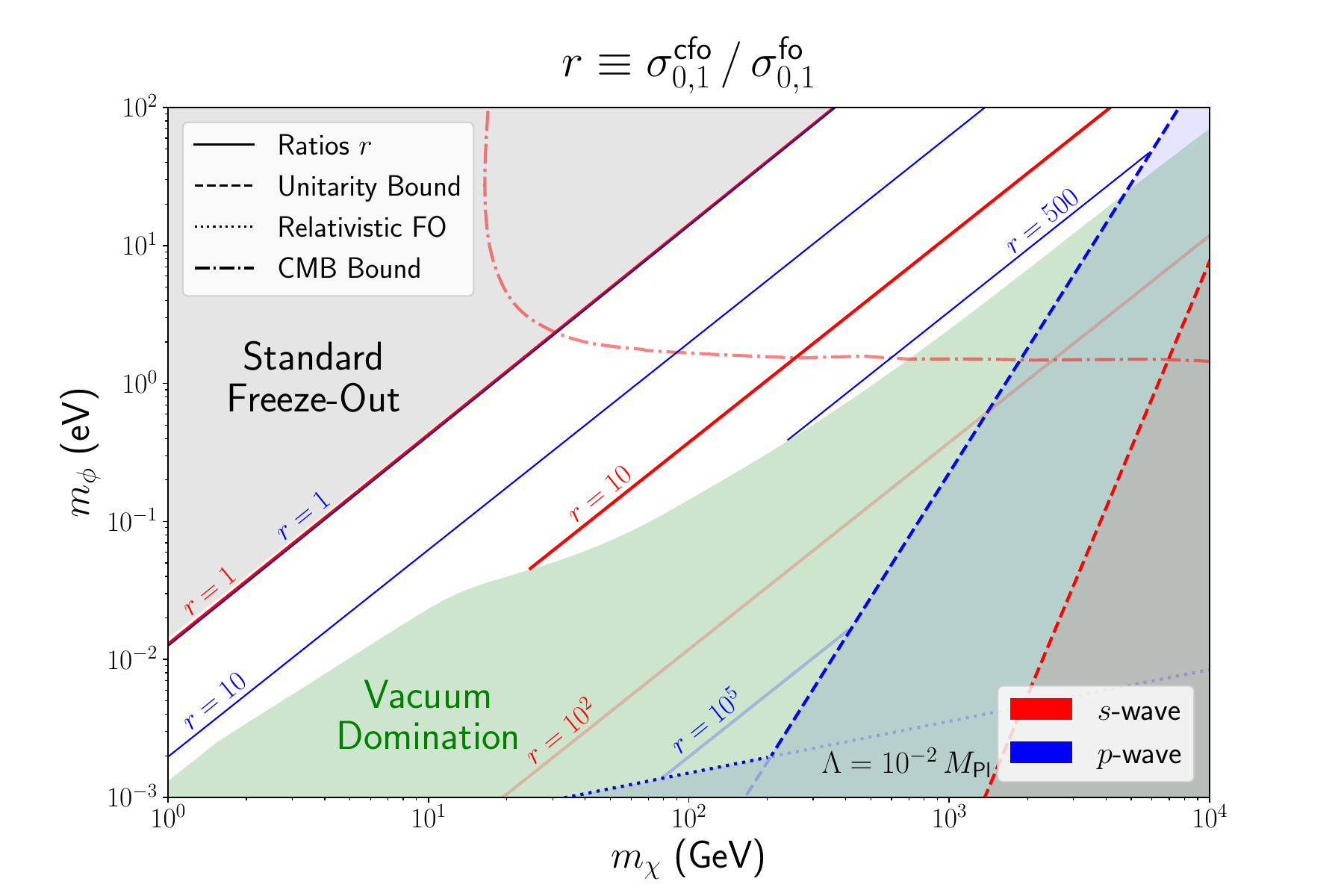}
\caption{\label{fig:boostsigma}
Coherent freeze-out mechanism in the \textbf{FOPT} regime, where DM consists solely of the WIMP. 
For illustration, we fix the cutoff scale to $\Lambda = 10^{-2} M_{\rm Pl}$. 
Contour lines show the ratio of WIMP annihilation cross sections in the coherent and standard freeze-out scenarios for 
$s$-wave (red solid) and $p$-wave (blue solid) processes, assuming both reproduce the observed relic abundance. The green shaded region denotes the parameter space where the vacuum energy dominates before tunneling occurs.
Regions below the red (blue) dotted curves are excluded because freeze-out would occur while the WIMP is still relativistic for $s$-wave ($p$-wave). 
Regions to the right of the red (blue) dashed curves violate unitarity for $s$-wave ($p$-wave). 
The gray shaded region marks $r=1$, where coherent and standard freeze-out coincide. 
The red dash-dotted line shows the CMB constraint on the $s$-wave cross section, excluding the region below it. 
No analogous CMB bound applies to $p$-wave DM due to velocity suppression.
}
\end{figure*}

Freeze-out is initiated when symmetry is restored. In the parameter space of interest, $\xcfotilde > x_{\rm osc}$, so after $\xcfotilde$, the ALP energy density redshifts as matter, $\varphi (x>\xcfotilde) = \varphi_* (\xcfotilde)\times(\xcfotilde/x)^{3/2}$. This increases the WIMP effective mass:
\begin{align}
\frac{\mchieff}{m_\chi} &= 1 - \frac{\varphi_*^2(\xcfotilde)}{2}\frac{\xcfotilde^3}{x^3} = 1 - \frac{\xcfotilde^3}{x^3} + c_1 \frac{\xcfotilde^2}{x^3}\;.\label{eq:mchieffovermchi}
\end{align}
The rapid approach to $\mchieff = m_\chi$ after $\xcfotilde$ drives 
$\chi$ out of equilibrium. This indicates the beginning of \textbf{coherent freeze-out}.

Using this expression in the Boltzmann equation and expanding $\sigma v = \sigma_0 + \sigma_1 v^2 + \cdots$, 
the coherent freeze-out temperature $\xcfo$ is found to be
\begin{align}
\xcfo &\approx \xcfotilde + \frac{1}{3}\log c+ \frac{7}{6}\log\left(\frac{1}{3}\log c\right),\label{eq:xcfo}
\end{align}
with
\begin{align}
c & \equiv \frac{27g_\chi}{2\pi^3} \sqrt{\frac{15}{2g_*}} m_\chi M_{\rm Pl} \left(\sigma_0 + 6 \sigma_1 /\xcfotilde\right) \xcfotilde^{-4}\;.
\end{align}
The resulting asymptotic yield is
\begin{align}
Y_{\chi,\infty}^{\rm cfo} \approx \frac{14}{9} \sqrt{\frac{45}{\pi g_{*}}} \frac{1}{m_\chi M_{\rm Pl}}\frac{\xcfo}{\sigma_0 + 21\sigma_1/(10 \xcfo)}\;.
\label{eq:Yinfcfo-maintext}
\end{align}
The derivations of (\ref{eq:xcfo})-(\ref{eq:Yinfcfo-maintext}) are provided in Appendix~\ref{app:WIMPdynamics}.
These analytical expressions are highly accurate in the small-$\kappa$ regime, where $\xcfo \approx \xcfotilde$; deviations from the exact numerical results remain below  ${\cal O}(1)$\% for $\xcfotilde \gtrsim  50$ (see Table~\ref{tab:compare}). The evolution of $Y_\chi$ with temperature is shown in Fig.~\ref{fig:abundance}.

At this point, it is interesting to compare the yield predicted in our mechanism with that in the standard freeze-out case:
\begin{align}
\frac{Y_{\chi,\infty}^{\rm cfo}}{Y_{\chi,\infty}^{\rm fo}} = \frac{14}{9}\frac{\xcfo}{x_{\rm fo}} \frac{\sigma_0^{\rm fo} + 3\sigma_1^{\rm fo}/x_{\rm fo}}{\sigma_0^{\rm cfo}+21 \sigma_1^{\rm cfo}/(10\xcfo)}\;,\label{eq:yieldratio}   
\end{align}
where $\sigma_i^{\rm cfo}$ and $\sigma_i^{\rm fo}$ denote the annihilation cross sections in the coherent and standard freeze-out cases, respectively, and $x_{\rm fo}\approx 25$ is the standard freeze-out temperature. For $\sigma_i^{\rm cfo}=\sigma_i^{\rm fo}$, we find an enhancement of the yield of order $\xcfo/x_{\rm fo}$ for $s$-wave and $\xcfo^2/x_{\rm fo}^2$ for $p$-wave annihilation. Since $\xcfo \sim \kappa^{-1/3} \gg 25$, the enhancement -- especially for $p$-wave -- is substantial.

If $\kappa$ is extremely small, freeze-out may occur when $\chi$ is still relativistic. For instance, if WIMPs annihilate via exchanging weak gauge bosons, for temperatures below the weak scale, we have $\langle \sigma v\rangle_{\rm eff} \sim T^2 G_F^2/x^\ell$, 
with $\ell =0$ ($\ell = 1$) for $s$-wave ($p$-wave), and $G_F$ is the Fermi constant. Requiring $n_\chi \langle \sigma v\rangle_{\rm eff}\sim H$ with $n_\chi \sim T^3$ gives the relativistic freeze-out temperatures:
\begin{align}
&
T_{\rm rfo}^s \sim G_F^{-2/3} M_{\rm Pl}^{-1/3} \sim {\rm MeV}\;,\\
&
T_{\rm rfo}^\text{$p$} \sim G_F^{-1/2} \left(\frac{m_\chi}{M_{\rm Pl}}\right)^{\frac{1}{4}} \sim 10~{\rm MeV} \left(\frac{m_\chi}{10~{\rm GeV}}\right)^{\frac{1}{4}}.
\end{align}
Relativistic freeze-out yields too large $\chi$ relic abundance, so viable models require $\xcfo < m_\chi/T_{\rm rfo}$.

The evolution of $\phi$ during freeze-out puts another upper bound on $\xcfo$. During phase transition, the field is first trapped in the symmetry-breaking vacuum and then undergoes non-adiabatic evolution, tunneling under the potential barrier. For sufficiently small $\kappa$, vacuum energy can dominate before tunneling occurs, leading to a new inflationary phase. Avoiding vacuum domination places a conservative bound on the latest possible onset of coherent freeze-out, $\xcfo \lesssim {\cal O}(500)$ (see Appendix~\ref{subapp:ALPdecay} for further discussion).

In the coherent freeze-out scenario, the WIMP annihilation cross section can exceed the standard value while still yielding the correct relic abundance. Fig.~\ref{fig:boostsigma} shows the ratio of the required cross sections in the two scenarios, assuming both reproduce the observed DM density, where we fix  $g_\chi=2$. The $r=1$ contour corresponds to $\kappa \approx 10^{-4}$. For $\kappa \gtrsim 10^{-4}$, the ALP symmetry is restored before $x_{\rm fo} \approx 25$, reducing to the standard freeze-out case (gray region), whereas for $\kappa \lesssim 10^{-4}$ the coherent effect becomes significant.

Cosmic microwave background (CMB) anisotropies place stringent limits on the WIMP $s$-wave annihilation cross section $\sigma_0$~\cite{Slatyer:2015jla,Slatyer:2015kla}. For $m_\chi \lesssim 30~\mathrm{GeV}$, the thermal value $\sigma_0^{\rm fo} \approx 2.2\times10^{-26}~{\rm cm^3/s}$ is excluded~\cite{Planck:2018vyg}, with the bound weakening at higher masses. In the coherent freeze-out scenario, the required $s$-wave cross section can exceed the standard value, extending the CMB constraint to heavier WIMPs. This bound appears as the red dash-dotted line in Fig.~\ref{fig:boostsigma}, and the region below it is excluded for $s$-wave-dominated annihilation. Even under this constraint, the $s$-wave cross section can still be enhanced by up to two orders of magnitude before reaching the unitarity limit~\cite{Griest:1989wd} (red dashed line),
and by a factor of about 30 before reaching the vacuum-domination limit (green shaded region).

In contrast, the enhancement needed for $p$-wave annihilation is much larger -- scaling as $\propto \xcfo^{2}/x_{\rm fo}^{2}$ rather than $\propto \xcfo/x_{\rm fo}$ -- and can reach values of order $10^{5}$ in Fig.~\ref{fig:boostsigma} before saturating unitarity and about $10^3$ before reaching the vacuum-domination limit. Consequently, $p$-wave-dominated WIMPs can have annihilation cross sections up to three orders of magnitude above the standard thermal value under the conservative assumption that vacuum domination does not occur while still yielding the observed relic abundance. This greatly enlarges the viable parameter space and may improve the prospects for the indirect detection of $p$-wave DM.


Finally, we comment on the ALP evolution after WIMP freeze-out. For $x>\xcfotilde$, the field oscillates around $\phi=0$, with the ALP energy density red-shifting as matter. An explicit estimate gives the ALP density after freeze-out, $\rho_\phi (T) \sim m_\chi T^3$, which, if stable, would exceed the observed DM density and overclose the Universe. 
Thus, $\phi$ must be unstable and should also decay rapidly into radiation after the onset of coherent oscillations to avoid significant dilution of the WIMP yield. The decay of $\phi$ is generally model-dependent and does not affect the parameter space shown in Fig.~\ref{fig:boostsigma}; for completeness, an explicit example realizing the rapid decay of $\phi$ is provided in Appendix~\ref{subapp:ALPdecay}.
We therefore conclude that, in the FOPT regime, DM is composed solely of the WIMP.

\subsection{Crossover regime}

\begin{figure*}[t!]
\centering
\includegraphics[width=1.0\textwidth]
{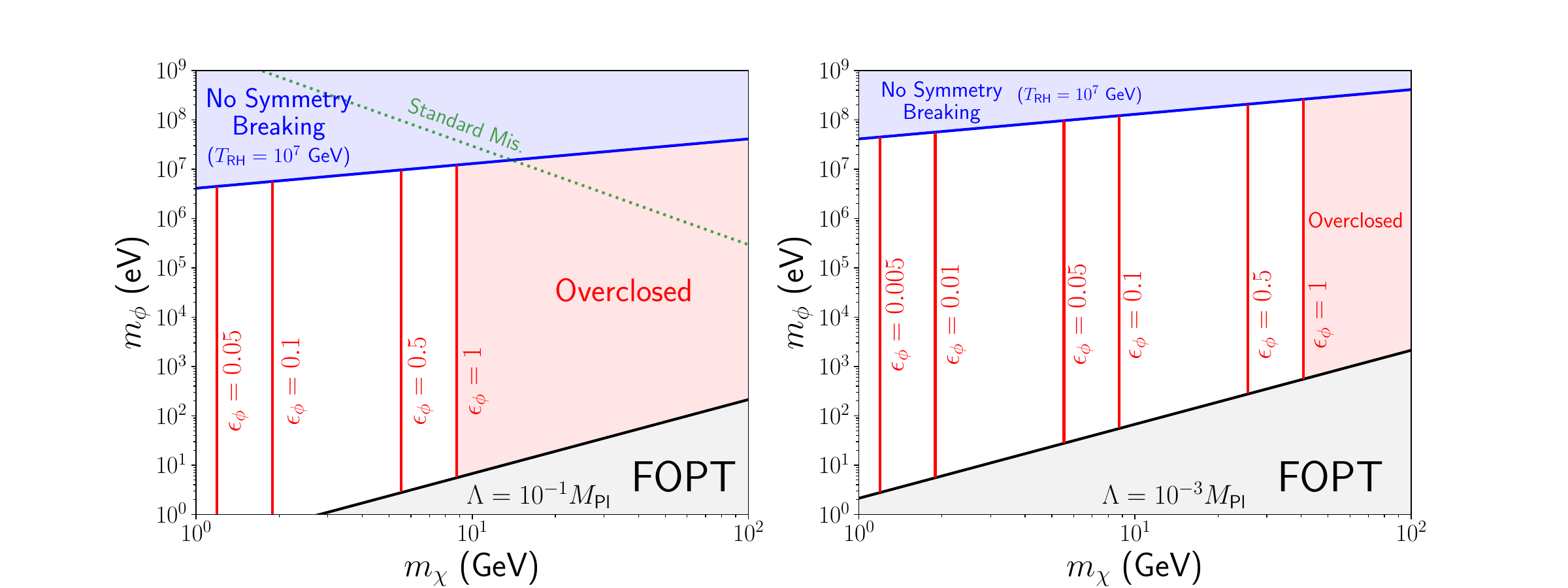}
\caption{\label{fig:MoneyPlot2}
Coherent freeze-out mechanism in the \textbf{crossover} regime, where DM may consist of both the
WIMP and the ALP. 
We fix the cutoff scale to $\Lambda = 10^{-1} M_{\rm Pl}$ (left) and $\Lambda = 10^{-3} M_{\rm Pl}$ (right). 
Red contours show the ALP fraction $\epsilon_\phi \equiv \Omega_\phi/\Omega_{\rm DM}$; 
regions with $\epsilon_\phi > 1$ overproduce ALP DM and are excluded. 
The area below the black line ($\kappa = \kappa_c$) corresponds to the FOPT regime, 
while regions above the blue line exhibit no symmetry breaking (relaxable with a higher reheating temperature). 
For comparison, the green dotted line shows the standard misalignment prediction of $\epsilon_\phi = 1$,  
using $\phi_i = \sqrt{m_\chi \Lambda}$ as the initial field displacement.
}
\end{figure*}

We now turn to the crossover regime, where the ALP evolution is nearly adiabatic.
In this regime, the symmetry of the ALP potential is restored before $x_{\rm fo} \approx 25$, so the ALP has little influence on the WIMP freeze-out. 
Instead, the thermal effect from the WIMP bath significantly alters the ALP evolution, producing a relic abundance that departs from the standard misalignment result. 
Consequently, the ALP can naturally serve as a second DM component in this regime.

Since the two sectors decouple while $\chi$ is still relativistic, we may take the limit $\gamma x \ll 1$ in the ALP EOM (\ref{eq:EOMvarphi}), which reduces to
\begin{align}
\varphi'' + \frac{2}{x} \varphi' -\eta\left(1-\kappa x^2\right)\varphi + \eta\,\varphi^3=0 \;.\label{eq:EOMrelativistic}
\end{align}
The symmetry is broken for $x < x_c \equiv 1/\sqrt{\kappa}$ and restored for $x > x_c$. 
As shown in Fig.~\ref{fig:solutions1st2nd}, the ALP is initially frozen by Hubble friction for $x \ll 1/\sqrt{\eta}$; 
for $1/\sqrt{\eta} \lesssim x \lesssim x_c$, it evolves toward the symmetry-breaking minimum at $\varphi_*=1$ and stabilizes there, erasing any dependence on its initial value; 
and for $x \gtrsim x_c$, the symmetry is restored and the field relaxes back toward $\varphi = 0$.

The adiabatic approximation holds when
\begin{align}
\eta \left|1-\kappa x^2 \right| \gg 1/x^2\;,   
\label{eq:adiabaticapproximation}
\end{align}
i.e., when the effective ALP mass dominates over Hubble friction. This condition fails near $x_c$, where the thermal and bare mass terms cancel. We define $x_1 = x_c + \delta x$ by requiring $\left|\eta\left(1-\kappa x_1^2\right)\right|\equiv 1/x_1^2$ so that adiabaticity is restored for $x > x_1$. In the limit $\eta \gg \kappa$ (which is always satisfied for the relevant parameter space), one finds $\delta x/x_c \approx \kappa/(2\eta)$, implying that the loss of adiabaticity is extremely brief.

For $x>x_1$, adiabatic evolution gives
\begin{align}
\mphieff (x) \phi^2(x)\,a^3(x) = \mphieff(x_1) \phi^2(x_1)\, a^3(x_1)\;, \label{eq:adiabaticinvariant}
\end{align}
where $a(x)$ is the cosmic scale factor.
The quantity in (\ref{eq:adiabaticinvariant}) is the adiabatic invariant, corresponding to the conservation of the comoving ALP number density.

At present, $x_0\equiv m_\chi/T_0$, the thermal mass is negligible, $\mphieff(x_0)=m_\phi$, so we obtain
\begin{align}
\frac{\phi^2(x_0)\,a^3(x_0)}{\phi^2(x_1) \,a^3(x_1)} = \frac{H(x_1)}{m_\phi} \approx \frac{H(x_c)}{m_\phi}  = \left(\frac{\kappa}{\eta}\right)^{\frac{1}{2}}\ll 1\;.\label{eq:phidrop}    
\end{align}
This corresponds to a rapid drop in the ALP field around $x_c$ during symmetry restoration, which agrees well with the numerical solution (see Fig.~\ref{fig:SolZoom}). The present-day energy density is therefore
\begin{align}
\rho_{\phi}(x_0) &= m_\phi^2 \phi^2(x_1) \left(\frac{\kappa}{\eta}\right)^{\frac{1}{2}}\left(\frac{x_1}{x_0}\right)^3 \frac{g_{*S}(x_0)}{g_{*S}(x_1)}\nonumber\\
&\approx   m_\phi^2 \phi^2(x_c) \left(\frac{\kappa}{\eta}\right)^{\frac{1}{2}}\left(\frac{x_c}{x_0}\right)^3 \frac{g_{*S}(x_0)}{g_{*S}(x_c)}\;, \label{eq:rhophi0}    
\end{align}
where in the second line we used  $x_1 \approx x_c$ for $\eta \gg \kappa$ and $g_{*S}$ is the entropy-weighted number of relativistic degrees of freedom. Substituting  $\phi(x_c) = \sqrt{m_\chi \Lambda}$ into (\ref{eq:rhophi0}) and using (\ref{eq:kappadef}) and (\ref{eq:etadef}), we arrive at the relic abundance:

\begin{align}
\Omega_\phi &= \frac{1.66\cdot 4g_{*S}(x_0) \sqrt{2g_\chi g_*(x_c)}}{3 g_{*S}(x_c)}\, \frac{m_\chi^{3/2} \Lambda^{1/2} T_0^{3}}{H_0^2 M_{\rm Pl}^3}\nonumber\\
& \approx 0.3 \left(\frac{m_\chi}{10~{\rm GeV}}\right)^{3/2} \left(\frac{\Lambda}{0.1M_{\rm Pl}}\right)^{1/2},
\end{align}
where to obtain the numerical value, we used $g_\chi=4$, $g_{*S}(x_c) = g_{*}(x_c)=106.75$, and $g_{*S}(x_0) = 3.93$.

We thus arrive at an ``\textbf{ALP miracle}": an ALP that couples quadratically to a 
weak-scale fermion with a Planck-suppressed interaction can naturally obtain 
the correct relic abundance. 
Importantly, this prediction is insensitive to both the initial ALP field value 
and the ALP mass, provided the latter lies in the crossover regime, 
${\rm eV} \lesssim m_\phi \lesssim {\rm MeV}$ 
(see the right panel of Fig.~\ref{fig:potential}).

In Fig.~\ref{fig:MoneyPlot2}, we show contour lines of the ALP fraction 
$\epsilon_\phi \equiv \Omega_\phi/\Omega_{\rm DM}$, where 
$\Omega_{\rm DM} \simeq 0.26$ is the observed DM density~\cite{Planck:2018vyg}. 
For comparison, the dotted line shows the standard misalignment prediction of 
$\epsilon_\phi = 1$, obtained by taking the initial displacement 
$\phi_i = \sqrt{m_\chi \Lambda}$.

\section{Conclusions}

In this Letter, we considered the cosmological dynamics of a WIMP-like fermion and an ALP-like scalar field. We found that even a tiny effective coupling between the WIMP and ALP fields can have a dramatic effect on their evolution due to coherent forward scattering. At high temperatures, spontaneous symmetry breaking by the ALP field may be induced. The dynamics of the subsequent phase transition determine the present-day densities of both WIMPs and ALPs, which may deviate by many orders of magnitude from the predictions of the standard WIMP freeze-out and ALP misalignment production mechanisms. Many of the conventional experimental targets for WIMP and ALP searches need to be reconsidered in this scenario. For example, the indirect detection of a $p$-wave WIMP could become more promising. In addition, part of the parameter space of our model predicts a strongly first-order phase transition associated with the WIMP freeze-out, potentially sourcing an observable gravitational-wave signal. The mass suppression during the coherent freeze-out phase also implies efficient momentum exchange between WIMPs and the thermal bath, suggesting that kinetic decoupling occurs no earlier than the phase transition, although its precise timing and the resulting impact on small-scale structure formation are model-dependent.
These possibilities will be explored in future work~\cite{Ferrante:2026}.

\begin{acknowledgments}
We thank Kai Bartnick, Cédric Delaunay, and Seung J. Lee for useful discussions.
This work is supported by the NSF grant PHY-2309456.
SF is partially supported by the Boochever Fellowship at Cornell University.
\end{acknowledgments}

\bibliography{ref}
\clearpage

\widetext
\begin{appendix}
\section{Order of the Phase Transition}
\label{app:orderPT}
The potential in Eq.~(\ref{eq:potential}) is controlled by a single parameter $\kappa$.
In this appendix, we determine how the order of the phase transition depends on $\kappa$.

Ginzburg-Landau theory provides a convenient framework for analyzing phase transitions in terms of an order parameter, which in our case is the ALP vacuum expectation value.
Near the critical point, the effective Lagrangian can be expanded as
\begin{align}
    \mathcal{L} = 
    \frac{1}{2}(\partial_{\mu}\phi)^{2}
    - a(T)\phi^{2} 
    - b(T)\phi^{4}
    - c(T)\phi^{6} + \cdots \; . 
\end{align}

The vacuum configuration $\langle\phi\rangle$ that minimizes the potential 
\begin{align}
    V(\phi) = 
    a(T)\phi^{2} + 
    b(T)\phi^{4} + 
    c(T)\phi^{6} + \cdots 
\end{align}
is controlled primarily by the sign of the quadratic coefficient $a(T)$:
\begin{itemize}
    \item $a > 0: \,\,  
    \langle\phi\rangle = 0$
    \,\, symmetric phase;
    \item $a < 0: \,\,  
    \langle\phi\rangle \neq 0$
    \,\, broken phase. 
\end{itemize}
The temperature at which $a(T)$ changes sign defines the critical point, $a(T_{c})=0$. 
The order of the transition is determined by the quartic coefficient evaluated at $T_c$:
\begin{itemize}
    \item $b(T_{c}) > 0:$ \,\,  
     crossover transition;
    \item $b(T_{c}) < 0:$ \,\,  
    first-order transition.
\end{itemize}

We can map the Ginzburg-Landau parameters onto the potential used in the main text, Eq.~(\ref{eq:potential}), by expanding it in powers of the normalized field $\varphi$.
This yields
\begin{align}
    a(x) &= \frac{1}{2}
    \bigg(
    \kappa - \frac{K_{1}(x)}{x}
    \bigg),\label{eq:ax} 
    \\
    b(x) &= \frac{1}{4}
    \bigg(
    \frac{3K_{1}(x)}{x} - K_{2}(x)
    \bigg).
\end{align}
This directly determines the temperature dependence of the quadratic and quartic terms relevant for diagnosing the order of the phase transition.

The critical value of $\kappa$, denoted $\kappa_{c}$, separates the first-order phase transition (FOPT) from the crossover regime. 
The boundary is obtained by setting the quartic coefficient to zero, $b(x)=0$, which can be solved numerically and gives
\begin{align}
x_{\text{tri}} \approx 1.33\;.    
\end{align}
We identify this as the \emph{tricritical temperature} -- the value of $x$ at which the nature of the transition changes. 
If the critical temperature $x_{c} \equiv m_\chi/T_{c}$ lies above (below) $x_{\text{tri}}$, the transition is first order (crossover).

At the tricritical point, both coefficients vanish,
\begin{align}
a(x_{\text{tri}})=0\;, \qquad b(x_{\text{tri}})=0\;,    
\end{align}
marking the boundary between the two regimes. 
Substituting $x_{\text{tri}}=1.33$ into $a(x)=0$ yields
\begin{align}
\kappa_{c} \approx 0.27\;.    
\end{align}
The resulting classification of FOPT versus crossover in the $(m_\chi,m_\phi)$ plane is shown by the red curves in the right panel of Fig.~\ref{fig:potential}. 

Whether symmetry breaking can occur in the early Universe is determined by the curvature of the ALP potential near the origin, given by the sign of Eq.~(\ref{eq:ax}). Since $K_{1}(x)/x$ is a monotonically decreasing function of $x$, symmetry breaking is possible only if it already occurs at the reheating temperature $T_{\rm RH}$. This requires
\begin{align}
\kappa < \frac{K_1(x_{\rm RH})}{x_{\rm RH}}\;,
\label{eq:condition-of-PT}
\end{align}
where $x_{\rm RH}\equiv m_\chi/T_{\rm RH}$. This condition is shown by the blue curves in the right panel of Fig.~\ref{fig:potential}; above these lines, no phase transition occurs at any time during cosmic evolution.

One can apply the same method to the full one-loop thermal potential in Eq.~(\ref{eq:Vfullnormalized}) and obtain a critical value $\kappa_c \approx 0.21$. This result shows that the approximate potential in Eq.~(\ref{eq:potential}) provides a good description of the qualitative regimes of the phase transition.

\section{Full Solutions of the ALP Equation of Motion}
\label{app:solution}
\begin{figure}[t!]
\centering
\includegraphics[width=3.2in]
{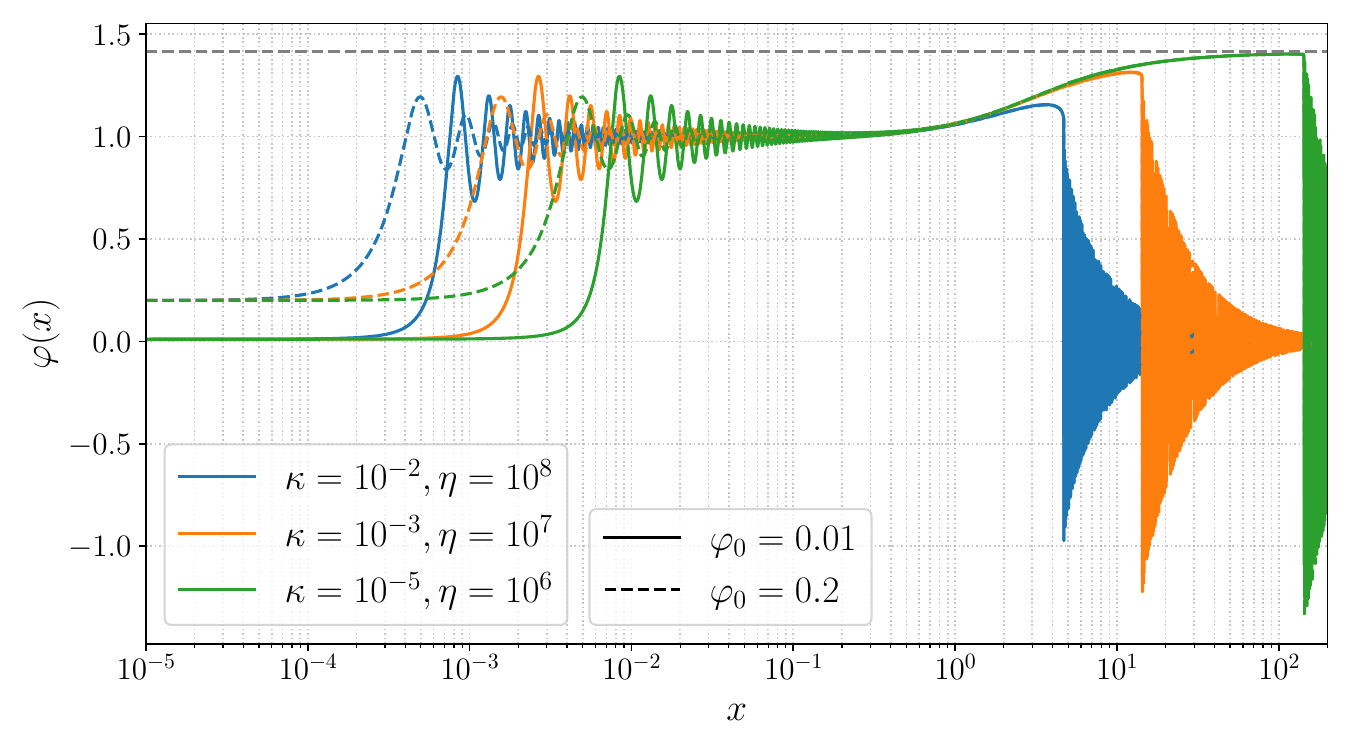}
\includegraphics[width=3.2in]
{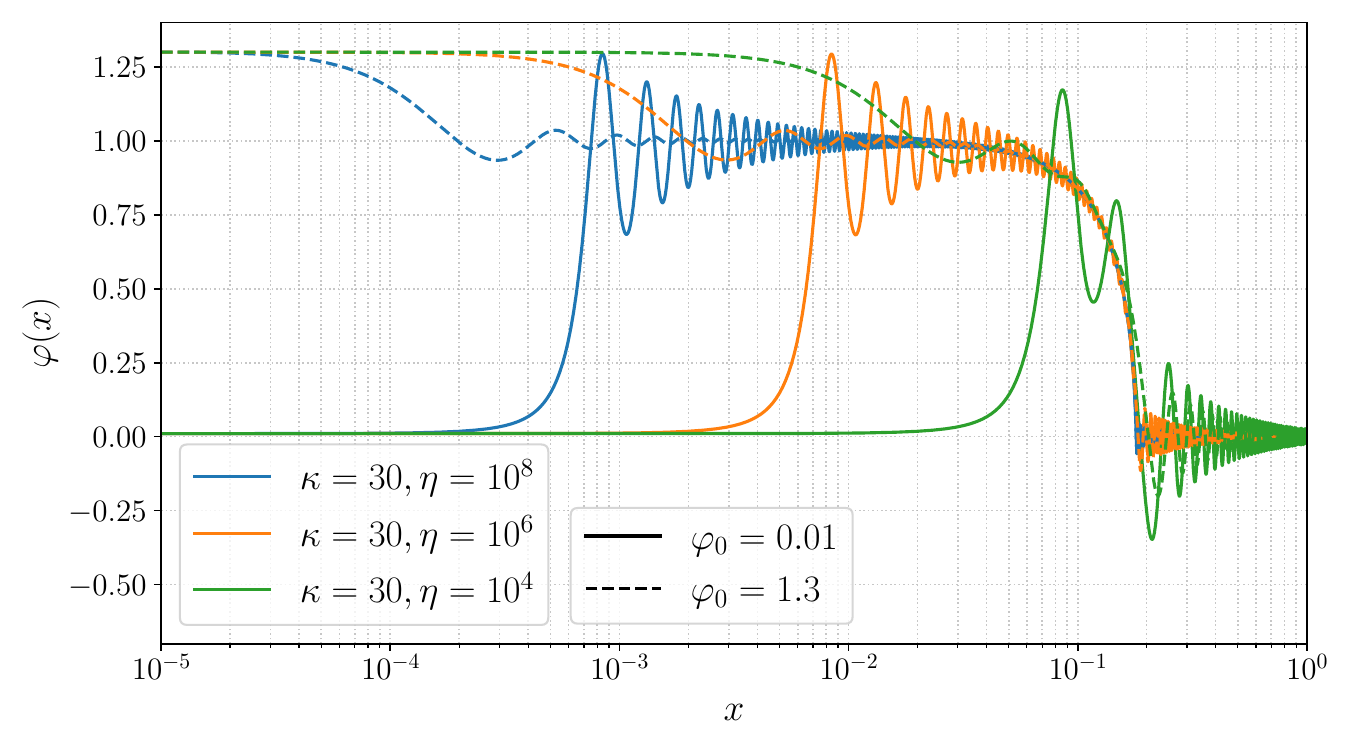}
\caption{\label{fig:solutions1st2nd}
Evolution of the ALP field in the first-order phase transition regime (left panel) and the crossover regime (right panel). 
Different colors correspond to different choices of $\kappa$ and $\eta$, while different line styles denote different initial field displacements.}
\end{figure}

The numerical solutions to the ALP equation of motion (\ref{eq:EOMvarphi}) are shown in Fig.~\ref{fig:solutions1st2nd} for both the FOPT and crossover regimes. 
Throughout, we assume that the ALP field has negligible initial velocity after inflation. 
To demonstrate the robustness of the evolution, we display trajectories with different initial displacements, $\varphi_0 = 10^{-2},\, 0.2,$ and $1.3$ (different line styles). 
As the figure illustrates, the qualitative dynamics of the ALP field are insensitive to its initial value.

In the upper panel of Fig.~\ref{fig:solutions1st2nd}, which shows the solution for $\varphi(x)$ in the FOPT regime ($\kappa < 0.27$), two distinct stages of oscillation are visible. 
First, the field oscillates about the symmetry-breaking minimum, located near $\varphi = 1$ and drifting toward $\sqrt{2}$ as $x$ increases. 
This behavior matches the black curves in Fig.~\ref{fig:potential} and also Fig.~\ref{fig:vacuum}, where the displaced minimum approaches $\sqrt{2}$ before disappearing at $\xcfotilde$. 
The onset of these early oscillations is controlled by $\eta$: larger $\eta$ (blue curve) leads to an earlier departure from Hubble friction. 
At $x=\xcfotilde$, the symmetry-breaking minimum vanishes, and the field rolls toward the origin, initiating a second set of oscillations that resemble standard misalignment. 
The onset of this phase is governed by $\kappa$, with smaller $\kappa$ (green curve) corresponding to a larger value of $\xcfotilde$.

In the lower panel of Fig.~\ref{fig:solutions1st2nd} we show the crossover regime ($\kappa > 0.27$), where the minimum of the potential smoothly evolves from $\varphi = 1$ to $\varphi = 0$. As in the FOPT case, the onset of the initial oscillations is controlled by $\eta$: larger $\eta$ causes the field to leave the Hubble-frozen regime earlier. 
Unlike the FOPT regime, however, the vacuum does not drift toward $\varphi = \sqrt{2}$; instead, it continuously interpolates from $\varphi=1$ to $\varphi=0$. 
The parameter $\kappa$ again determines when symmetry restoration occurs and thus when the second set of oscillations begins.

\begin{figure}[t!]
\centering
\includegraphics[width=5in]
{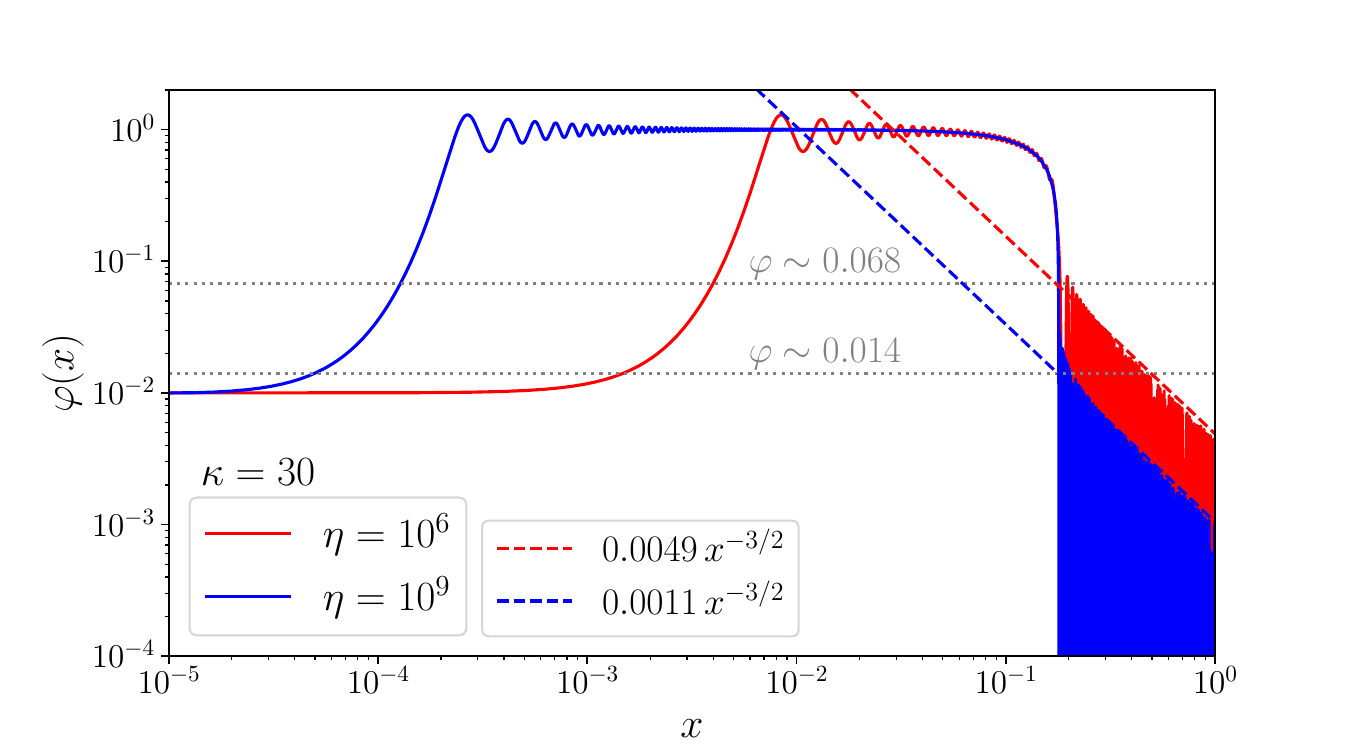}
\caption{\label{fig:SolZoom}
Zoomed-in view of the ALP evolution for $\eta = 10^{6}$ (red) and $\eta = 10^{9}$ (blue) in the crossover regime with fixed $\kappa = 30$. 
Dashed lines indicate the matter-like scaling behavior, with normalization chosen to match the corresponding ALP trajectories.}

\end{figure}

A key feature of the crossover regime is the rapid drop of the field value during symmetry restoration, a direct consequence of the nearly adiabatic evolution predicted by Eq.~(\ref{eq:phidrop}). 
This behavior underlies the ``ALP miracle" and is numerically confirmed in Fig.~\ref{fig:SolZoom}. 
For $\kappa = 30$ and $\eta = 10^{6}$ ($10^{9}$), Eq.~(\ref{eq:phidrop}) predicts a drop of order $(\kappa/\eta)^{1/4} \approx 0.074$ (0.013) near $x_c = 1/\sqrt{\kappa} \approx 0.18$. 
Just before the drop, Fig.~\ref{fig:SolZoom} shows $\varphi \approx 1$; the gray dotted lines at $\varphi \approx 0.068$ and $0.014$ mark the post-drop values for $\eta = 10^{6}$ and $10^{9}$, respectively, prior to the onset of matter-like redshifting. 
The agreement with the analytical prediction is especially good for $\eta = 10^{9}$, as the adiabatic approximation becomes increasingly accurate at larger $\eta$, consistent with Eq.~(\ref{eq:adiabaticapproximation}). This validates the analytical treatment used in the main text for the range of $\eta$ relevant to the ALP-miracle window in Fig.~\ref{fig:MoneyPlot2}, namely $\eta \sim 10^{15}-10^{19}$.

\section{Coherent Freeze-Out Dynamics in the Strong First-Order Phase Transition Region}
\label{app:WIMPdynamics}
In this section, to gain analytical insight, we restrict ourselves to the approximate potential in Eq.~(\ref{eq:potential}) for the analytical derivation of the coherent freeze-out dynamics. We begin by studying the evolution of the ALP vacuum before symmetry restoration, which determines the onset of coherent freeze-out. We then compute the coherent freeze-out temperature $\xcfo$ and the resulting WIMP relic yield in this scenario. Finally, we discuss the ALP evolution after symmetry restoration and the associated cosmological constraints.

A more complete treatment using the full potential in Eq.~(\ref{eq:Vfullnormalized}) does not modify the analytical expressions derived below for the coherent freeze-out temperature and the WIMP yield.

\subsection{Evolution of the ALP Vacuum}
\label{app:vacuum}
\begin{figure*}[t!]
\centering
\includegraphics[width=3.5in]
{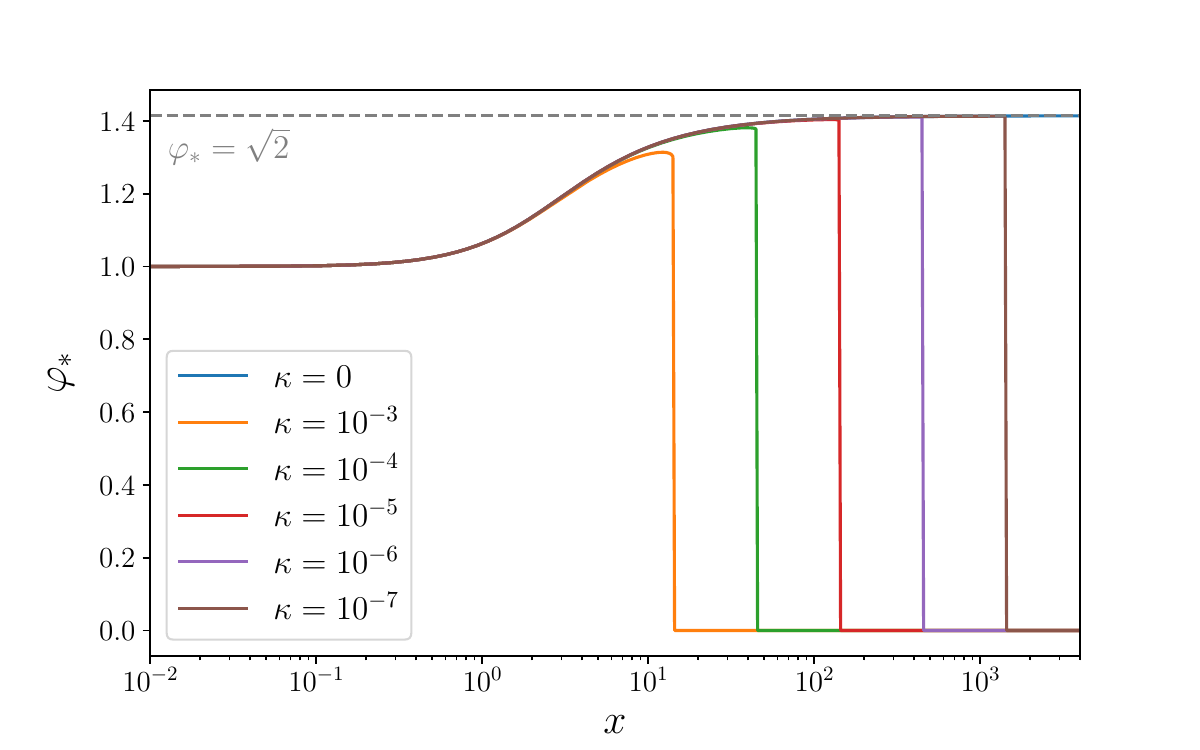}
\includegraphics[width=3.5in]
{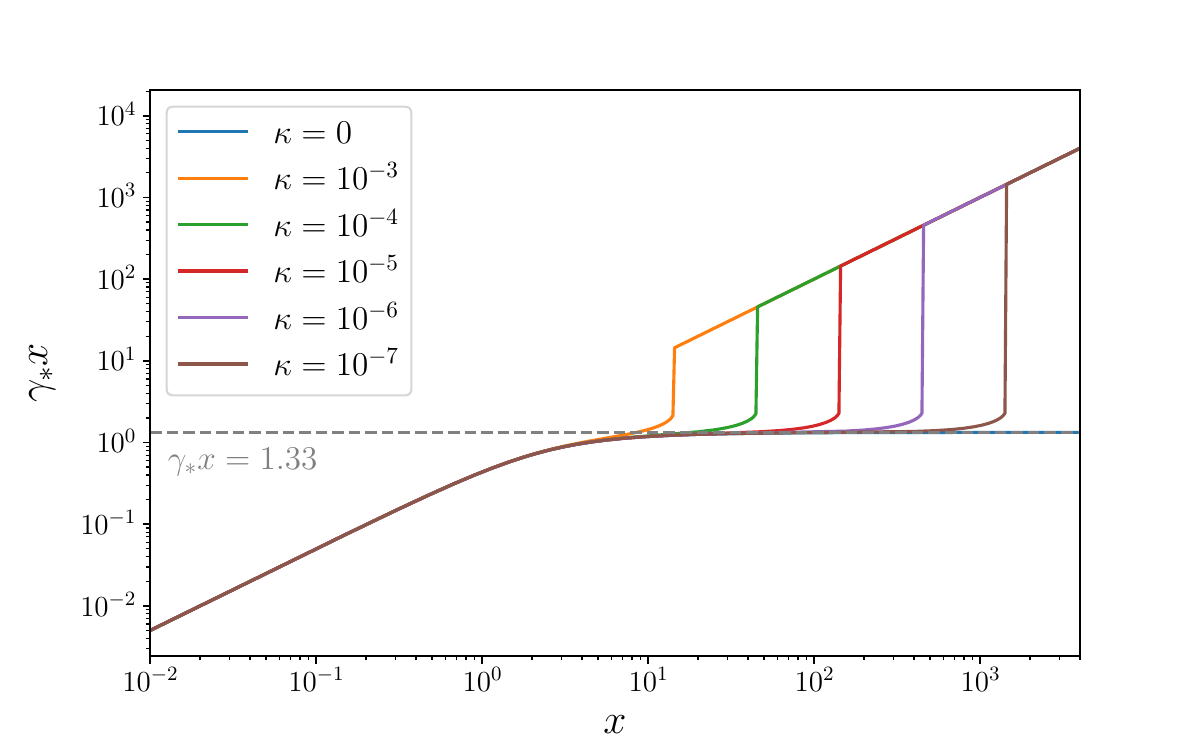}
\caption{\label{fig:vacuum}The evolution of the symmetry-breaking vacuum $\varphi_*$ (left panel) and the boost factor $\gamma_* \equiv |1-\varphi_*^2/2|$ (right panel) as a function of the inverse temperature $x\equiv m_\chi/T$. The ratio $\mchieff/T=\gamma_* x$ remains smaller than ${\cal O}(1)$ until $x$ reaches $\xcfotilde$ (corresponding to the sudden drop of $\varphi_*$ and the sharp increase of $\gamma_* x$ in the plots), after which the coherent freeze-out begins.}
\end{figure*}

When the symmetry is broken, the displaced vacuum $\varphi_* \neq 0$ is determined by the nonzero solution of
\begin{align}
\frac{\partial{\cal V}(\varphi,x)}{\partial \varphi} &= \kappa \varphi - \frac{\varphi^3}{2}\left(1-\varphi^2/2\right)^2 K_0(x\left(1-\varphi^2/2\right))-\frac{\varphi}{2x}\left(2-3\varphi^2+\varphi^4\right) K_1(x(1-\varphi^2/2)) = 0\;,   \label{eq:vacuum}
\end{align}
where ${\cal V}$ is given by (\ref{eq:potential}). This equation can be solved numerically for any fixed $\kappa$, as shown in Fig.~\ref{fig:vacuum}. We are particularly interested in the small-$\kappa$ limit relevant for a strong FOPT. Neglecting the $\kappa$ term, the condition simplifies to
\begin{align}
\frac{\gamma_*x\,K_0(\gamma_*x)}{K_1(\gamma_*x)} = \frac{2(\varphi_*^2-1)}{\varphi_*^2}\;,\label{eq:varphistarsmallkappa}    
\end{align}
where $\gamma_* \equiv 1-\varphi_*^2/2$ is the boost factor evaluated at the symmetry-breaking vacuum. The solution of $\varphi_* (x)$ is shown in the left panel of Fig.~\ref{fig:vacuum}: one finds $\varphi_*(x) \to 1$ for $x\ll 1$ and $\varphi_*(x) \to \sqrt{2}$ for $x\gg 1$ (consistent with Fig.~\ref{fig:solutions1st2nd}). 

An interesting feature emerges in the large-$x$ limit: the product of the boost factor and inverse temperature approaches a constant (see the right panel Fig.~\ref{fig:vacuum}): 
\begin{align}
\gamma_* x \approx 1.33\;.
\label{eq:gammatimesx}
\end{align}
(This numerical value corresponds to the root of the transcendental equation $c_1 K_0(c_1) = K_1(c_1)$ where $c_1\equiv \gamma_*x$.)

This result has a striking impact on WIMP freeze-out. 
If $\kappa$ is sufficiently small, the evolution of the symmetry-breaking vacuum described by Eq.~(\ref{eq:varphistarsmallkappa}) remains a good approximation even at $x \gg 1$. 
In this regime, the effective WIMP mass satisfies $\mchieff/T = \gamma_* x \approx 1.33$, implying that the WIMP remains in thermal equilibrium regardless of how large the bare ratio $x \equiv m_\chi/T$ becomes. 
Freeze-out can therefore occur only after symmetry is restored (corresponding to the jump in Fig.~\ref{fig:vacuum}), when the ALP rolls back to the origin and $\mchieff \to m_\chi$. 
As a result, the WIMP remains relativistic for much longer than in the standard scenario, and the coherent freeze-out temperature $T_{\rm cfo}$ can be significantly lower, allowing 
$x_{\rm cfo} \equiv m_\chi/T_{\rm cfo} \gg \mathcal{O}(25)$. This delayed onset of Boltzmann suppression arises from the medium-induced suppression of the effective WIMP mass and annihilation rate, leading to less redshifting between freeze-out and the present day and thereby enhancing the relic abundance relative to the standard case.
In the next subsection, we quantify this effect.

\subsection{WIMP Relic Abundance}
The Boltzmann equation that governs $\chi$ freeze-out dynamics is given by
\begin{align}
\frac{{\rm d}n_\chi}{{\rm d}t} + 3 Hn_\chi = - \langle \sigma v\rangle_{\rm eff} \left(n_\chi^2-n_{\chi,{\rm eq}}^2\right), \label{eq:Boltzmann}
\end{align}
where $n_\chi$ is the number density of $\chi$ and $\langle \sigma v\rangle_{\rm eff}$ is the effective cross section of two $\chi$ particles annihilating to SM particles. In the medium of the ALP background, the effective mass of the WIMP is shifted from $m_\chi$ to $\mchieff = \gamma(\varphi)\,m_\chi$, where $\gamma(\varphi)=|1-\varphi^2/2|$ is the field-dependent boost factor. As a result, the effective annihilation cross section is also modified.

We assume that $\chi$ couples to SM species through a weak-scale interaction; therefore, we have $\langle \sigma v\rangle_{\rm eff} \sim 1/(\mchieff^2\,x^\ell)$ for $\mchieff \gtrsim m_{\rm M}$ and $\langle \sigma v\rangle_{\rm eff} \sim \mchieff^2/(m_{\rm M}^4 \, x^\ell)$ for $\mchieff \lesssim m_{\rm M}$, where $m_{\rm M}$ denotes the mass of the mediator that mediates the interaction between WIMP and SM particles, typically of the weak scale, and $\ell = 0$ ($\ell =1$) for $s$-wave ($p$-wave). Putting them together, we can write $\langle \sigma v\rangle_{\rm eff} \sim \min \left\{1/(\gamma^2 m_\chi^2 x^\ell),\, \gamma^2 m_\chi^2/(m_{\rm M}^4 x^\ell) \right\}$. 

We are interested in the effect of the ALP field on WIMP freeze-out dynamics. In the crossover region, the symmetry of the ALP potential is restored earlier, typically at $x<x_{\rm fo}\approx 25$, so the effect on freeze-out is negligible. By contrast, the effect is significant in the strong FOPT region with $\kappa \ll 1$, as we discuss below.

From the right panel of Fig.~\ref{fig:vacuum}, it can be seen that the boost factor scales as $\gamma \sim  1/x$ for $ 1 \ll x \lesssim \xcfotilde$, which suppresses $\mchieff$ to keep it well below $m_{\rm M}$. Therefore, we can factor out the dependence of the annihilation cross section on the ALP field value:
\begin{align}
\langle \sigma v\rangle_{\rm eff} = \gamma^2 \langle\sigma v \rangle \;,      
\end{align}
where $\langle\sigma v \rangle \sim m_\chi^2/(m_{\rm M}^4 x^\ell)$ is the \emph{bare} annihilation cross section that does not depend on $\varphi$. In the non-relativistic limit, we can expand the bare cross section over the velocity: $\sigma v=\sigma_0+\sigma_1 v^2 + \cdots$, where $\sigma_0$ ($\sigma_1$) denotes the $s$-wave ($p$-wave) cross section.
After the thermal average, it becomes $\langle \sigma v \rangle = \sigma_0 + 6\,\sigma_1/x + \cdots$. We neglect terms more suppressed than $p$-wave.
The equilibrium number density of $\chi$ is also affected by $\varphi$ through the effective mass of $\chi$ (assuming a Maxwell-Boltzmann distribution of the $\chi$ bath):
\begin{align}
n_{\chi,{\rm eq}} = \frac{g_\chi}{2\pi^2} \mchieff^2 T K_2(\mchieff/T)\;.    
\end{align}

Introducing the yield $Y_\chi\equiv n_\chi/s$ with $s$ the entropy density, the Boltzmann equation (\ref{eq:Boltzmann}) is reduced to
\begin{align}
Y_\chi'(x) = -\frac{\lambda\gamma^2}{x^2}\left(1+\frac{b}{x}\right) \left[Y_\chi^2(x)-Y_{\chi,{\rm eq}}^2(x)\right],
\label{eq:BoltzmannY}
\end{align}
where $b  \equiv 6\,\sigma_1/\sigma_0$, and
\begin{align}
\lambda &\equiv \sqrt{\frac{\pi}{45}} \frac{g_{*S}}{\sqrt{g_*}} m_\chi M_{\rm Pl} \sigma_0\;,\label{eq:lambdadef}\\
Y_{\chi,{\rm eq}}(x) &\equiv \frac{n_{\chi,{\rm eq}}}{s} = \frac{45 g_\chi}{4\pi^4 g_{*S}} \gamma^2 x^2 K_2(\gamma x)\;,\label{eq:Yeq}
\end{align}
with $g_*$ and $g_{*S}$ denoting the number of relativistic degrees of freedom relevant to energy and entropy, respectively. Note that the equilibrium yield, $Y_{\chi,{\rm eq}}$, depends essentially on $\gamma x =\mchieff/T$ rather than $m_\chi/T$.

\begin{figure}[t!]
\centering
\includegraphics[width=3.5in]
{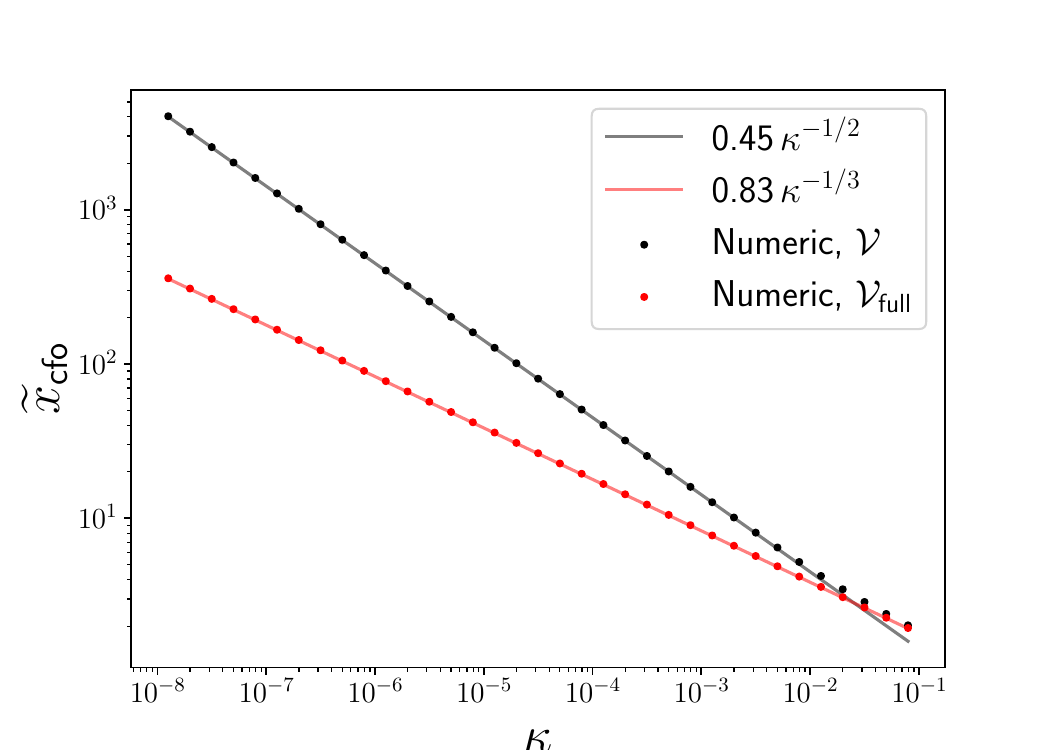}
\caption{\label{fig:xcfovskappa}
$\xcfotilde$ as a function of $\kappa$. 
The dots show the numerical solution of Eq.~(\ref{eq:xcfotildedef}), while the solid lines denote the fitting functions in Eqs.~(\ref{eq:xcfotildequad})-(\ref{eq:xcfotildefull}) using the approximate and full thermal potentials, respectively.
The agreement is excellent in the small-$\kappa$ regime.}
\end{figure}

In the broken phase, the ALP field is located at the symmetry-breaking vacuum $\varphi_* \neq 0$, reducing the effective WIMP mass. In the small-$\kappa$ limit, $\gamma x =\mchieff/T  \approx 1.33$ remains a constant (see Eq.~(\ref{eq:gammatimesx})) regardless of how large the value of $x=m_\chi/T$ is. This allows WIMP to stay in equilibrium even at $x \gg {\cal O}(25)$. The local minimum at $\varphi_*$ disappears at $\xcfotilde$, defined by 
\begin{align}
 \frac{\partial^2{\cal V}(\varphi,\xcfotilde)}{\partial \varphi^2} \Bigg|_{\varphi=\varphi_*} = 0\;. \label{eq:xcfotildedef}   
\end{align}
The solution of (\ref{eq:xcfotildedef}) is shown in Fig.~\ref{fig:xcfovskappa}. For $\kappa \ll 1$, it can be fitted by a simple power law
\begin{align}
\xcfotilde &\approx 0.45\, \kappa^{-1/2}\;,\quad \text{for ${\cal V}$ = Eq.~(\ref{eq:potential})}\;,\label{eq:xcfotildequad}\\
\xcfotilde &\approx 0.83 \,\kappa^{-1/3}\;,\quad \text{for ${\cal V}$ = Eq.~(\ref{eq:Vfullnormalized})}\;.\label{eq:xcfotildefull}
\end{align}

We observe that $\xcfotilde$ exhibits a slightly different scaling with respect to
$\kappa$ depending on whether one uses the approximate MB potential in Eq.~(\ref{eq:potential}) or the full potential in Eq.~(\ref{eq:Vfullnormalized}). The derivation of Eq.~(\ref{eq:xcfotildefull}) is provided in Appendix~\ref{app:fullthermal}. For a fixed value of $\xcfotilde$, however, this difference does not affect the subsequent discussion of the freeze-out dynamics. It only becomes relevant when translating a given value of $\xcfotilde$ to $\kappa$, which depends on the underlying physical parameters. In Figs.~\ref{fig:abundance} and \ref{fig:boostsigma} of the main text as well as Table~\ref{tab:compare}, we always use Eq.~(\ref{eq:xcfotildefull}) to provide more accurate numerical results.

At $x>\xcfotilde$, the ALP rolls back to the origin with a matter-like equation of state (one can check that $m_\phi>H$ is always satisfied at $\xcfotilde$ for the parameter space of interest). Averaging over a period, it gives
\begin{align}
\varphi (x>\xcfotilde) = \left(\xcfotilde/x\right)^{3/2} \varphi_*\;.
\label{eq:phidecrease}
\end{align}
Therefore, the boost factor at $x > \xcfotilde$ scales as
\begin{align}
\gamma &= 1- \frac{\varphi_*^2(\xcfotilde)}{2} \frac{\xcfotilde^3}{x^3}= 1- \frac{\xcfotilde^3}{x^3}+ c_1 \frac{\xcfotilde^2}{x^3}\approx 1- \frac{\xcfotilde^3}{x^3}\;,\label{eq:boostfactor}  
\end{align}
where the approximation in the last step holds for $1\ll \xcfotilde < x < \xcfo$. We see from (\ref{eq:boostfactor}) that after $\xcfotilde$, the effective mass of the WIMP tends to its bare mass rapidly as $\gamma \to 1$, which triggers the coherent freeze-out.
Substituting (\ref{eq:boostfactor}) into the Boltzmann equation (\ref{eq:BoltzmannY}) determines the coherent freeze-out temperature $\xcfo$. The numerical solution of (\ref{eq:BoltzmannY}) is shown in Fig.~\ref{fig:abundance}.
In the following, we show how to calculate $\xcfo
$ and the final yield analytically.

Our approximate analytical method is inspired by the method in \cite{Kolb:1990vq} that applies to the standard freeze-out case. Before freeze-out ($x < \xcfo$), the yield can track closely with the equilibrium value, $Y_\chi \approx Y_{\chi,{\rm eq}}$, so (\ref{eq:BoltzmannY}) becomes 
\begin{align}
Y_\chi - Y_{\chi,{\rm eq}} \approx -\frac{x^2}{2\lambda \gamma^2 (1+b/x)} \frac{Y_{\chi,{\rm eq}}'}{Y_{\chi,{\rm eq}}}\;. \label{eq:YdiffersYeq}
\end{align}
Using the non-relativistic expansion that satisfies at $\gamma x \gg 1$, we have $Y_{\chi,{\rm eq}}\approx a (\gamma x)^{3/2} e^{-\gamma x}$, where $a\equiv 45 g_\chi/(4\sqrt{2} \pi^{7/2}g_{*S})$ is roughly a constant during freeze-out. Using (\ref{eq:boostfactor}), we obtain
\begin{align}
\frac{Y_{\chi,{\rm eq}}'}{Y_{\chi,{\rm eq}}} \approx - \left(1+2 \frac{\xcfotilde^3}{x^3}\right), \quad \text{for $x > \xcfotilde \gg 1$}\;.\label{eq:YeqprimeoverYeq}  
\end{align}
We define the freeze-out temperature $\xcfo$ as the time after which $Y_\chi$ fails to track $Y_{\chi,{\rm eq}}$,
\begin{align}
Y_\chi(\xcfo) - Y_{\chi,{\rm eq}}(\xcfo) \equiv Y_{\chi,{\rm eq}}(\xcfo)\;.
\label{eq:xcfodef}
\end{align}
Combining Eqs.~(\ref{eq:boostfactor})-(\ref{eq:xcfodef}), we obtain 
\begin{align}
&\xcfo^{1/2}\left(1+2\frac{\xcfotilde^3}{\xcfo^3}\right)\,e^{\left(1-\xcfotilde^3/\xcfo^3\right)\,\xcfo} = 2a\lambda \left(1+\frac{b}{\xcfo}\right) \left(1-\frac{\xcfotilde^3}{\xcfo^3}\right)^{7/2}.   
\end{align}
Because freeze-out occurs rapidly after $\xcfotilde$, we have $\Delta x \equiv \xcfo -\xcfotilde \ll \xcfotilde$. Note that the non-relativistic approximation of $Y_{\chi,{\rm eq}}$ used above requires $3\Delta x \gg 1$.
Up to the leading order of $\Delta x/\xcfotilde$, the above equation is reduced to
\begin{align}
(\Delta x)^{-7/2} e^{3 \Delta x} = 18 \sqrt{3}\,a\lambda\left(1+b/\xcfotilde\right)\xcfotilde^{-4}\;.   
\end{align}
This can be solved iteratively, which leads to the freeze-out temperature
\begin{align}
\xcfo &\approx \xcfotilde + \frac{1}{3}\log\left(18 \sqrt{3}\,a\lambda\,\left(1+b/\xcfotilde\right) \xcfotilde^{-4} \right) + \frac{7}{6}\log\left[\frac{1}{3}\log\left(18 \sqrt{3}\,a\lambda\left(1+b/\xcfotilde\right)\xcfotilde^{-4} \right)\right].\label{eq:xcfoprediction}  
\end{align}
The analytical formula in (\ref{eq:xcfoprediction}) is accurate in the limit of $\kappa \ll 1$ (corresponding to $\xcfotilde \gg 1$). Note that the value of $\Delta x$ decreases as $\kappa$ decreases.
For sufficiently small $\kappa$, the condition $3 \Delta x \gg 1$ breaks down, so that the above non-relativistic expansion of $Y_{\chi,{\rm eq}}$ fails. In this case, it is a good approximation to take $\xcfo \approx \xcfotilde$.

Next, we move on to calculate the yield. After freeze-out, $Y_{\chi,{\rm eq}}$ is negligible compared to $Y_\chi$, so the Boltzmann equation (\ref{eq:BoltzmannY}) is reduced to
\begin{align}
\frac{{\rm d}Y_\chi}{Y_\chi^2} = -\frac{\lambda}{x^2}\left(1+\frac{b}{x}\right)\left(1-\frac{\xcfotilde^3}{x^3}\right)^2 {\rm d}x\;.
\end{align}
Integrating from $x_{\rm cfo}$ to infinity at both sides leads to
\begin{align}
\frac{1}{Y_{\chi,\infty}} - \frac{1}{Y_\chi(\xcfo)} &= \frac{\lambda}{\xcfo}\left(1-\frac{1}{2} \frac{\xcfotilde^3}{\xcfo^3} + \frac{1}{7} \frac{\xcfotilde^6}{\xcfo^6} \right)\nonumber\\
&+ \frac{b \lambda}{2\xcfo^2}\left(1-\frac{4}{5} \frac{\xcfotilde^3}{\xcfo^3} + \frac{1}{4} \frac{\xcfotilde^6}{\xcfo^6} \right)
\end{align}
where the first (second) line denotes the contribution from $s$-wave ($p$-wave). Up to the leading order of $\Delta x/\xcfotilde$, the final yield in the coherent freeze-out scenario is given by
\begin{align}
Y_{\chi,\infty}^{\rm cfo} &\approx \frac{14}{9} \frac{\xcfo}{\lambda} \frac{1}{1+7b/(20 \xcfo)}\nonumber\\
& = \frac{14}{9} \sqrt{\frac{45 g_*}{\pi g_{*S}^2}} \frac{1}{m_\chi M_{\rm Pl}}\frac{\xcfo}{\sigma_0 + 21\sigma_1/(10 \xcfo)}\;,
\label{eq:Yinfcfo}
\end{align}
where in the second line we have used the relation $b=6\,\sigma_1/\sigma_0$ and (\ref{eq:lambdadef}).

\renewcommand\arraystretch{1.3}
\begin{table}[t!]
\centering
\begin{tabular}{c | c || c | c | c }
\hline
& & \,Analytic\, & \,Numeric\, & \,Error\,  \\ 
\hline\hline
\multirow{2}{*}{$\kappa = 10^{-4}$}
& $\xcfo$ & 26 & 31 & 16\%  \\ 
& $Y_{\infty} (\times 10^{-13}$) & 7.5 & 6.1 & 23\%  \\ 
\hline      
\multirow{2}{*}{$\kappa = 10^{-5}$}
& $\xcfo$ & 45.3 & 47.2 & 4.0\%  \\ 
& $Y_{\infty} (\times 10^{-12}$) & 1.32 & 1.13 & 17\%  \\ 
\hline
\multirow{2}{*}{$\kappa = 10^{-6}$}
& $\xcfo$ & 88.5 & 89.4 & 1.0\%  \\ 
& $Y_{\infty} (\times 10^{-12}$) & 2.57 & 2.40 & 7.1\%  \\ 
\hline
\multirow{2}{*}{$\kappa = 10^{-7}$}
& $\xcfo$ & 183.0 & 183.7 & 0.4\%  \\ 
& $Y_{\infty} (\times 10^{-12}$) & 5.31 & 5.19 & 2.3\%  \\ 
\hline
\multirow{2}{*}{$\kappa = 10^{-8}$}
& $\xcfo$ & 387.9 & 388.5 & 0.2\%  \\ 
& $Y_{\infty} (\times 10^{-11}$) & 1.125 & 1.123 & 0.2\%  \\ 
\hline
\end{tabular}
\caption{Comparison between the approximate analytical formulae (\ref{eq:xcfoprediction}) and (\ref{eq:Yinfcfo}) with the numerical results. For the analytical prediction, we also use Eq.~(\ref{eq:xcfotildefull}) to translate $\kappa$ into $\xcfotilde$.
The benchmark parameters used are $g_\chi=2$, $m_\chi = 1~\mathrm{TeV}$ and $\sigma_0 = 2.2\times10^{-26}~\mathrm{cm^3/s}$ (for $s$-wave annihilation).}
\label{tab:compare}
\end{table}
\renewcommand\arraystretch{1.0}

In Table~\ref{tab:compare}, we compare the analytical expressions for the freeze-out 
temperature~(\ref{eq:xcfoprediction}) and final yield~(\ref{eq:Yinfcfo}) with the 
results obtained from a full numerical solution of the Boltzmann 
equation~(\ref{eq:BoltzmannY}). 
As expected, the analytical formulae become increasingly accurate in the small-$\kappa$ 
(large-$\xcfo$) regime. 
Indeed, Table~\ref{tab:compare} shows that they already achieve $\sim {\cal O}(1)\%$ accuracy 
for $\xcfo \gtrsim 50$, corresponding to $\kappa \lesssim 10^{-5}$.

As a reference point, the standard freeze-out mechanism predicts a final yield~\cite{Cirelli:2024ssz}
\begin{align}
Y_{\chi,\infty}^{\rm fo} =  \sqrt{\frac{45 g_*}{\pi g_{*S}^2}} \frac{1}{m_\chi M_{\rm Pl}} \frac{x_{\rm fo}}{\sigma_0+3\sigma_1/x_{\rm fo}}\;,\label{eq:Yinffo}   
\end{align}
with $x_{\rm fo} \approx 25$. Assuming the same annihilation cross section, the coherent freeze-out mechanism predicts a final yield larger than the standard freeze-out by
\begin{align}
\frac{Y_{\chi,\infty}^{\rm cfo}}{Y_{\chi,\infty}^{\rm fo}} = \frac{14}{9}\frac{\xcfo}{x_{\rm fo}} \frac{\sigma_0 + 3\sigma_1/x_{\rm fo}}{\sigma_0+21 \sigma_1/(10\xcfo)}\;.\label{eq:yieldratio} 
\end{align}

For $s$-wave-dominated annihilation, the relic abundance is therefore enhanced by a factor of $\xcfo/x_{\rm fo}$. If $\chi$ is a Majorana fermion, the $s$-wave annihilation to a light SM fermion-antifermion pair ($\chi\chi \to f\bar{f}$) is helicity suppressed by $m_f^2/m_\chi^2$~\cite{Goldberg:1983nd};  therefore, for $m_\chi < m_W$ the annihilation of Majorana WIMP to SM particles is typically $p$-wave dominated. Taking $\sigma_0 \ll \sigma_1$ in (\ref{eq:yieldratio}), we obtain the enhancement factor for $p$-wave:
\begin{align}
\frac{Y_{\chi,\infty}^{\rm cfo}}{Y_{\chi,\infty}^{\rm fo}}\Bigg|_\text{$p$-wave} = \frac{20}{9} \frac{\xcfo^2}{x_{\rm fo}^2}\;,\label{eq:p-waveratio}
\end{align}
which is much more significant than the $s$-wave case. This result is remarkable: it allows $\sigma_1$ to exceed the standard thermal value by a factor of $(\xcfo/x_{\rm fo})^{2}$ while still producing the correct relic abundance, thereby opening the possibility of the indirect detection of $p$-wave DM.

We note that in the regime where the effective WIMP mass, $\mchieff \sim T$, becomes very small due to large $\xcfo$, some SM annihilation channels may become kinematically forbidden. The precise annihilation rate in this regime (i.e., the values of $\sigma_0$ and $\sigma_1$) therefore depends on the specific mediator structure and available final states of the underlying model that describes WIMP-SM interactions. In models where the dominant annihilation channel becomes kinematically inaccessible, the total annihilation rate may be further suppressed, leading to an additional enhancement of the relic abundance.
In this work we parametrize the annihilation cross section only by its velocity scaling (e.g., $s$-wave or $p$-wave) in order to illustrate the general impact of coherent freeze-out. In the yield estimate shown in Fig.~\ref{fig:abundance} and Table~\ref{tab:compare}, we fix $\sigma_0$ and $\sigma_1$ to benchmark values that reproduce the observed relic abundance in the standard freeze-out scenario.
Detailed threshold effects may modify the quantitative yield in specific models, but they do not affect the qualitative behavior of the coherent freeze-out mechanism discussed here.

Finally, we briefly comment on kinetic equilibrium during the coherent freeze-out phase. Since $m_{\chi,{\rm eff}}\sim T$ in this regime, elastic scattering with the thermal bath can efficiently exchange momentum ($\Delta p/p\sim\mathcal{O}(1)$ per collision). As a result, kinetic equilibrium is expected to be maintained throughout the coherent freeze-out stage, and kinetic decoupling cannot occur before $x_{\rm cfo}$. Whether kinetic decoupling takes place immediately after the phase transition or at a later time depends on the mediator responsible for WIMP-SM elastic scattering and is therefore model dependent. A dedicated analysis of the resulting kinetic-decoupling temperature and its implications for structure formation is beyond the scope of this work.

\subsection{ALP Evolution in the Coherent Freeze-out Scenario}
\label{subapp:ALPdecay}
For sufficiently small values of $\kappa$, the ALP can remain trapped in the false vacuum for too long, eventually causing its vacuum energy to dominate the total energy density of the Universe. To quantify this, we define the ratio between the ALP energy density evaluated at $\varphi=\varphi_*$ and the radiation energy density:
\begin{align}
\alpha(x) \equiv \frac{\rho_\phi}{\rho_{\rm rad}} \approx \frac{\frac{g_\chi}{2\pi^2}m_\chi^4 \kappa}{\frac{\pi^2}{30}g_*T^4}= \frac{15 g_\chi}{\pi^4 g_*} \kappa x^4\;,  
\end{align}
where the approximation holds in the region $x\gg 1$.
For a fixed $\kappa$, $\alpha$ increases as $x^4$. We define $x_{\rm d}$ as the value of $x$ at which the vacuum energy begins to dominate over radiation:
\begin{align}
\alpha(x_{\rm d})\equiv 1 \quad \Rightarrow \quad x_{\rm d} &= \left(\frac{\pi^4 g_*}{15 g_\chi}\right)^{1/4} \kappa^{-1/4}\;.\label{eq:xd}
\end{align}

On the other hand, the temperature at which the false vacuum disappears scales as $\xcfotilde \sim \kappa^{-1/3}$ (see Eq.~(\ref{eq:xcfotildefull})), which increases more rapidly with decreasing $\kappa$ than $x_{\rm d}$. Moreover, since $\Lambda \gg m_\chi$, tunneling through the potential barrier is highly suppressed until $x \approx \xcfotilde$.
As a result, for sufficiently small $\kappa$, the vacuum energy can dominate while the ALP is still trapped in the false vacuum, thereby triggering a new inflationary phase of the Universe. If this inflationary period persists long enough, it may significantly dilute the WIMP yield. The WIMP yield after inflation depends on both the number of e-folds and the details of reheating. A complete analysis of this inflationary phase is beyond the scope of this work and is left for future study.

In this work, we aim to identify the viable parameter space while avoiding significant dilution. To this end, we impose the conservative requirement that tunneling occur before vacuum domination, namely $\xcfotilde < x_{\rm d}$ (or equivalently, $\alpha(\xcfotilde )<1$). This condition sets a conservative lower bound on $\kappa$:
\begin{align}
\kappa \gtrsim \left(\frac{0.83^4 \cdot 15 g_\chi}{\pi^4 g_*}\right)^3\approx 4.3 \times 10^{-9} \left(\frac{g_\chi}{2}\right)^3 \left(\frac{90}{g_*}\right)^3.    
\end{align}
This, in turn, implies an upper bound on $\xcfotilde$:
\begin{align}
\xcfotilde \lesssim \frac{\pi^4 g_*}{0.83^3\cdot 15 g_\chi}\approx 511 \left(\frac{2}{g_\chi}\right) \left(\frac{g_*}{90}\right),\label{eq:xcfoupperlimit}   
\end{align}
where $g_*$ is evaluated at $x=\xcfotilde$.
Therefore, avoiding vacuum domination places a general constraint on the latest possible onset of coherent freeze-out, as shown in Eq.~(\ref{eq:xcfoupperlimit}). This bound is translated into the green shaded region in Fig.~\ref{fig:boostsigma}, and is stronger for lighter WIMP due to the smaller value of $g_*(\xcfotilde)$. 

Using Eq.~(\ref{eq:yieldratio}) and $x_{\rm fo}\approx 25$, the yield enhancement for $s$-wave and $p$-wave annihilation is roughly bounded by a factor of 30 and a factor of $10^3$, respectively,
\begin{align}
\frac{Y_{\chi,\infty}^{\rm cfo}}{Y_{\chi,\infty}^{\rm fo}}\Bigg|_\text{$s$-wave} &= \frac{14}{9} \frac{\xcfo}{x_{\rm fo}} \approx 30 \left(\frac{\xcfo}{500}\right),\\
\frac{Y_{\chi,\infty}^{\rm cfo}}{Y_{\chi,\infty}^{\rm fo}}\Bigg|_\text{$p$-wave} &= \frac{20}{9} \frac{\xcfo^2}{x_{\rm fo}^2} \approx 900 \left(\frac{\xcfo}{500}\right)^2.    
\end{align}

After $\xcfotilde$, the ALP field rolls down to the origin and subsequently redshifts as matter, marking the onset of coherent freeze-out. The ALP must decay at a later time to avoid overclosing the Universe. We require that the ALP energy be transferred to the SM radiation before Big Bang Nucleosynthesis (BBN) in order to evade the stringent cosmological bounds on $\Delta N_{\rm eff}$. In addition, the ALP should decay sufficiently rapidly after $\xcfotilde$ to prevent further dilution of the WIMP yield due to entropy injection into the SM bath. The decay of the ALP is generally model-dependent and does not affect the parameter space presented in the main text, which is based on the effective coupling between the WIMP and the ALP. Below, we provide an explicit example in which the ALP decays sufficiently rapidly after $\xcfotilde$, while still maintaining a sufficiently small coupling to avoid thermalizing the ALP in the early Universe.
\vspace{0.2cm}

We consider a Chern-Simons coupling between the ALP and a dark photon $A$,
\begin{align}
{\cal L}_{\rm dark} \supset -\frac{\phi}{4f_\phi} F_{\mu\nu}\tilde{F}^{\mu\nu}\;,\label{eq:darkphoton} 
\end{align}
where $f_\phi$ is the decay constant of $\phi$, and $F$ and $\tilde{F}$ denote the field strength of the dark photon and its dual, respectively.  We assume that $A$ is  massless at high temperatures and acquires a mass $m_{A}$ at $T\lesssim \Lambda_D$, where $\Lambda_D$ denotes the dynamical scale associated with dark photon mass generation. For $T \gtrsim \Lambda_D$, the dark photon is massless and the kinetic mixing with the SM photon is physically ineffective.
We take $ {\cal O}({\rm MeV})< \Lambda_D < m_\chi/\xcfotilde$, which ensures that the dark photon remains massless during the energy transfer epoch around $\xcfotilde$, while becoming massive sufficiently early to decay into the SM bath before BBN.

The coupling in Eq.~(\ref{eq:darkphoton}) is well known to induce exponential production of the massless gauge field, analogous to the preheating mechanism (see, e.g., \cite{Adshead:2015pva}). It therefore provides an efficient channel to rapidly transfer the ALP energy once the field begins coherent oscillations.
The EOM of $A$ in momentum space is 
\begin{align}
\ddot{A}_k^{\pm} + \left(k^2\mp k\frac{\dot{\phi}}{f_\phi}\right) A_k^{\pm} =0\;,\label{eq:darkphotoEOM}    
\end{align}
where $k$ is the comoving wavenumber. In writing Eq.~(\ref{eq:darkphotoEOM}), we have neglected the Hubble expansion, which is justified as long as $f_\phi \ll M_{\rm Pl}$. Modes in the range $0<k<|\dot{\phi}|/f_\phi$ undergo tachyonic growth, with the fastest-growing mode located at $k_* \sim |\dot{\phi}|/(2f_\phi)$. The corresponding maximal instantaneous growth rate is therefore $\mu_* \sim k_*$. Consequently, the gauge-field energy density grows exponentially as $\rho_A \propto \exp(2\int\mu_* {\rm d}t)$.
For a coherently oscillating field, one has $|\dot{\phi}| \sim \sqrt{\rho_\phi} \sim \sqrt{\alpha \rho_{\rm rad}}$, with $\alpha \lesssim 1$ evaluated at $x=\xcfotilde$. (To maximally delay WIMP freeze-out, we typically consider $\xcfotilde \sim x_{\rm d}$, implying $\alpha \sim 1$.) The characteristic timescale for the ALP to transfer its energy is then $\mu_*^{-1} \sim 2f_\phi/\sqrt{\alpha \rho_{\rm rad}}$. Comparing this with the Hubble timescale, we obtain
\begin{align}
\frac{\mu_*^{-1}}{H^{-1}} \sim \frac{5.8}{\sqrt{\alpha}} \frac{f_\phi}{M_{\rm Pl}}\;.   
\end{align}
As long as $f_\phi \lesssim 10^{-2} M_{\rm Pl}$, the ALP can transfer most of its energy into dark photons within one Hubble time, independently of the ALP mass in the range of interest. Furthermore, the Hubble expansion rate is suppressed by a factor of $f_\phi/M_{\rm Pl}$ compared to the characteristic growth rate $\mu_*$, thereby justifying the neglect of Hubble expansion in Eq.~(\ref{eq:darkphotoEOM}). 

This non-perturbative energy transfer process is parametrically much faster than the perturbative decay of the ALP. Indeed, one can compare the characteristic energy deposition rate in this mechanism, $\Gamma_{\rm preheat}\sim \mu_* \sim |\dot{\phi}|/(2f_\phi)$, with the perturbative decay rate of $\phi$ to two dark photons, $\Gamma_{\rm perturb} (\phi \to 2A) = m_\phi^3/(64\pi f_\phi^2)$:
\begin{align}
\frac{\Gamma_{\rm preheat}}{\Gamma_{\rm perturb}}\sim \frac{32\pi f_\phi |\dot{\phi}|}{m_\phi^3} \sim 32 \pi \sqrt{\alpha \frac{\pi^2}{30}g_* (\tilde{T}_{\rm cfo})}\, \frac{f_\phi \tilde{T}_{\rm cfo}^2}{m_\phi^3} \sim 10^{45} \left(\frac{{\rm eV}}{m_\phi}\right)^3 \left(\frac{f_\phi}{10^{16}~{\rm GeV}}\right)\left(\frac{\tilde{T}_{\rm cfo}}{{\rm GeV}}\right)^2,   
\end{align}
where $\tilde{T}_{\rm cfo}\equiv m_\chi/\xcfotilde$ denotes the temperature at which $\phi$ starts coherent oscillations around the origin. Therefore, the perturbative decay rate is completely negligible compared to the energy transfer rate associated with this preheating-like process. 

The preheating production of dark photons enters a nonlinear regime once the back-reaction of the produced gauge field on the ALP becomes significant, at which point the exponential amplification is effectively shut off. To estimate this effect, we include the coupling in Eq.~(\ref{eq:darkphoton}) in the ALP equation of motion:
\begin{align}
\ddot{\phi} + 3H \dot{\phi} + m_\phi^2 \phi = \frac{1}{f_\phi} \langle {\bf E} \cdot {\bf B} \rangle\;,\label{eq:ALPEOMsouceterm}    
\end{align}
where ${\bf E}$ and ${\bf B}$ denote the dark electric and magnetic fields, respectively, and $\langle \cdots \rangle$ denotes the ensemble average. The back-reaction becomes important when the source term on the right-hand side of Eq.~(\ref{eq:ALPEOMsouceterm}) is comparable to the mass term, i.e., when  $\left|\langle {\bf E} \cdot {\bf B} \rangle_{\rm br}\right| \sim f_\phi m_\phi^2 \phi_{\rm br}$, where $\phi_{\rm br}$ denotes the ALP field value at the onset of back-reaction. Using energy conservation, one finds $\phi_{\rm br}/\phi_0 \sim (\sqrt{z^2+4}-z)/2$, where $z\equiv f_\phi/\phi_0$, and $\phi_0 \sim \sqrt{m_\chi \Lambda}$ is the initial oscillation amplitude at $\xcfotilde$. In the region of $z\gg 1$, this simplifies to $\phi_{\rm br}/\phi_0 \sim 1/z$, implying that the remaining ALP energy fraction is suppressed as $\rho_{\phi, {\rm br}}/\rho_{\phi,0} \sim 1/z^2 \sim\phi_0^2/f_\phi^2$. Therefore, for sufficiently large $f_\phi$, most of the ALP energy is transferred before the back-reaction becomes relevant.

Combining these results, the viable parameter region in which the ALP avoids significant back-reaction during preheating while still transferring its energy efficiently within one Hubble time is
\begin{align}
   \sqrt{m_\chi \Lambda} \ll f_\phi \ll 10^{-2} M_{\rm Pl}\;. \label{eq:fphirange}
\end{align}
In the main text, we take $\Lambda \sim 10^{-2} M_{\rm Pl}$ and $m_\chi$ at the electroweak scale. The condition in Eq.~(\ref{eq:fphirange}) is therefore easily satisfied over a broad range of $f_\phi$, for any ALP mass within the region of interest.

If the time required for the ALP to release its energy into radiation is short compared to the Hubble timescale, the ALP deposits most of its energy almost instantaneously at $\xcfotilde$. In this case, the resulting dilution of the WIMP yield is mild, at most of ${\cal O}(1)$:
\begin{align}
Y_\chi \to \frac{Y_\chi}{\left(1+\alpha(\xcfotilde)\right)^{3/4}} > \frac{Y_\chi}{2^{3/4}} \approx 0.6\,Y_\chi\;,     
\end{align}
where we have used the condition $\alpha(\xcfotilde)<1$, as required by Eq.~(\ref{eq:xcfoupperlimit}).

After the ALP releases its energy, the dark photon remains massless until the temperature drops below $\Lambda_D$, at which point kinetic mixing with the SM photon is turned on. If $m_A \gtrsim  {\cal O}(30)\,{\rm MeV}$, the dark photon can efficiently transfer its energy to the SM bath through the decay channel $A \to e^+e^-$ well before BBN over a wide range of experimentally allowed kinetic mixing values, thereby avoiding the cosmological constraints on $\Delta N_{\rm eff}$.

\section{Full One-loop Thermal Potential} 
\label{app:fullthermal}

In the main text, we introduced two simplifying assumptions to gain analytical insight into the coherent freeze-out mechanism:
(i) we neglected quantum-statistical effects by using the classical Maxwell-Boltzmann (MB) distribution instead of the Fermi-Dirac (FD) distribution for the WIMP thermal bath, and
(ii) we ignored higher-order corrections in 
$\varphi$, retaining only the quadratic term in the one-loop thermal potential. (After including the back-reaction, the resulting potential also contains a term quartic in $\varphi$, ensuring it remains bounded from below.)
In this appendix, we show that incorporating both effects does not alter the qualitative picture presented in the main text.

\begin{figure}[t]
\centering
\includegraphics[width=3.4in]
{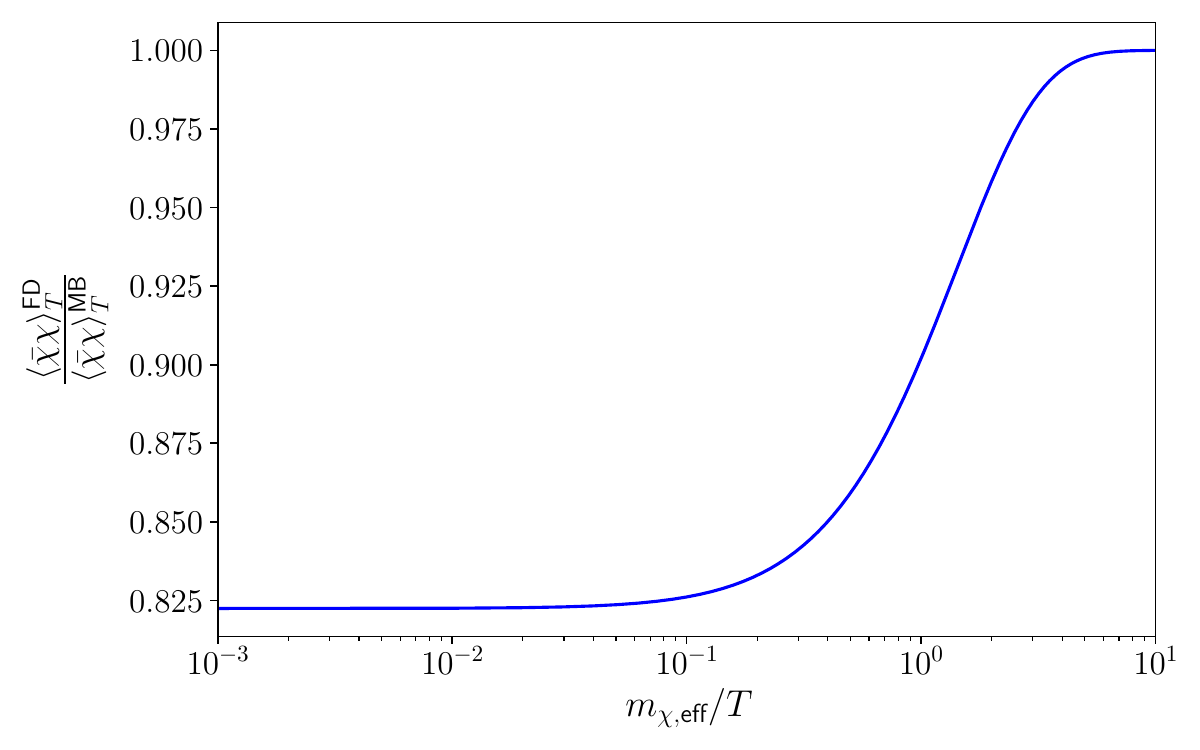}
\caption{\label{fig:MBFDCheck} Ratio of the Fermi-Dirac to Maxwell-Boltzmann contributions to the ALP thermal mass as a function of $\mchieff/T$. The ratio approaches $0.82$ in the relativistic limit and $1$ in the non-relativistic limit.}
\end{figure}

First, we compare the ALP thermal-mass contribution $m_\phi^2 \to m_\phi^2 -\langle \bar{\chi}\chi \rangle_T/\Lambda$ obtained using the MB and FD distributions. Substituting $f_\chi^{\rm MB} = e^{-E_\veck/T}$ and $f_\chi^{\rm FD} = 1/(e^{E_\veck/T}+1)$ into Eq.~(\ref{eq:chibarchi}) gives
\begin{align}
 \langle \bar{\chi} \chi\rangle_{T}^{\rm MB} &= \frac{g_\chi \mchieff }{2\pi^2} \int_{\mchieff}^{\infty} {\rm d}E\,\left(E^2-\mchieff^2\right)^{1/2} e^{-E/T} = \frac{g_\chi}{2\pi^2} \mchieff^2 T K_1(\mchieff/T)\;,\\
\langle \bar{\chi} \chi\rangle_{T}^{\rm FD} &= \frac{g_\chi \mchieff }{2\pi^2} \int_{\mchieff}^{\infty} {\rm d}E\,\left(E^2-\mchieff^2\right)^{1/2} \frac{1}{e^{E/T}+1}\;.
\end{align}
The FD expression does not admit a simple closed form. In the relativistic limit, $T \gg \mchieff$, both reduce to
\begin{align}
 \langle \bar{\chi} \chi\rangle_{T}^{\rm MB} & \approx \frac{g_\chi}{2\pi^2} \mchieff T^2\;\label{eq:MBrel}\\
 \langle \bar{\chi} \chi\rangle_{T}^{\rm FD} & \approx \frac{g_\chi}{24} \mchieff T^2\;.\label{eq:FDrel}
\end{align}
Thus, the two results differ by a factor of $\pi^2/(12) \approx 0.82$ in the relativistic limit and converge in the non-relativistic limit (see Fig.~\ref{fig:MBFDCheck}). Using the MB distribution instead of the FD distribution therefore induces an error of less than 20\%.

\begin{figure*}[t!]
\centering
\includegraphics[width=3.5in]
{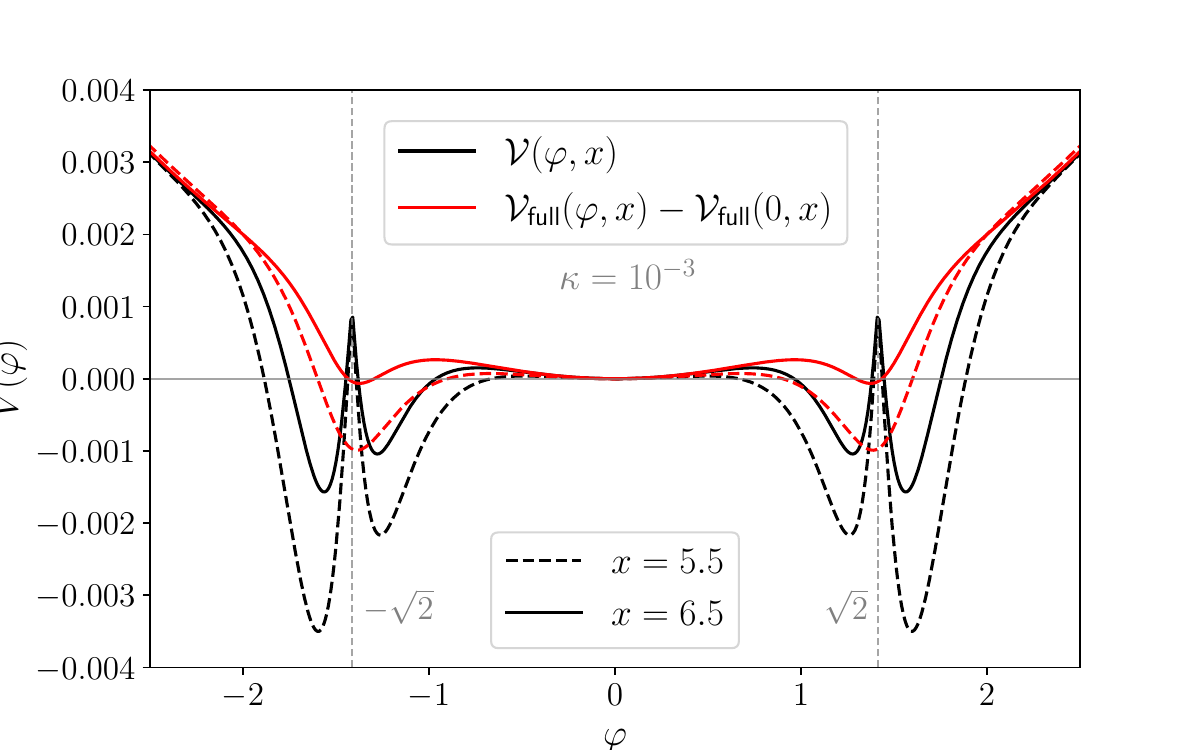}
\includegraphics[width=3.5in]
{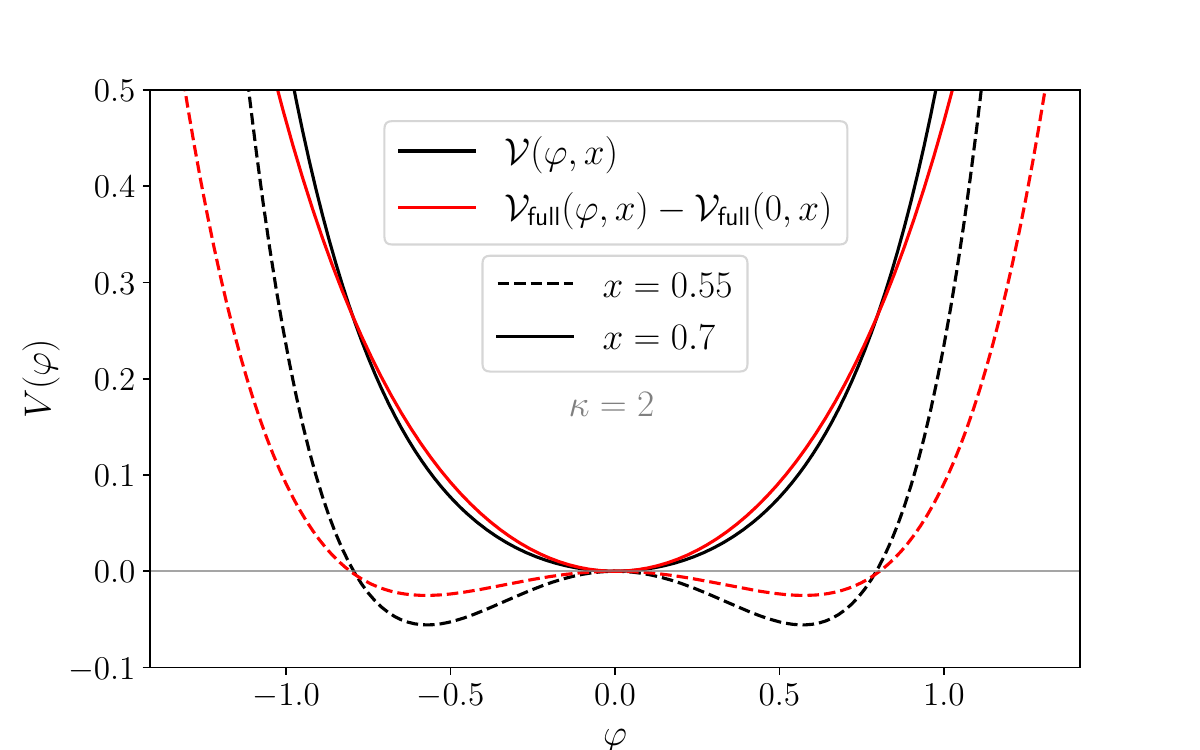}
\caption{\label{fig:ThermalPotential}
Comparison between the full one-loop thermal potential ${\cal V}_{\rm full}$ (red) in Eq.~(\ref{eq:Vfullnormalized}) and the approximate potential ${\cal V}$ (black) used in the main text in Eq.~(\ref{eq:potential}) in the FOPT regime ($\kappa = 10^{-3}$, left panel) and the crossover regime ($\kappa=2$, right panel), shown for several values of $x$. For clarity, the field-independent part of ${\cal V}_{\rm full}$ has been subtracted. The cusp at $\varphi = \pm\sqrt{2}$ (vertical dashed line) marks $\mchieff = 0$, beyond which the approximate potential breaks down and higher-order terms in $\varphi$ become important.}
\end{figure*}

Next, we consider corrections from higher-order terms in $\varphi$. The potential used in the main text is only valid for $\varphi \lesssim {\cal O}(1)$; for $\varphi \gg {\cal O}(1)$, one must include contributions from higher powers in $\varphi$. Given the coupling in (\ref{eq:Leff}), the full one-loop thermal potential of $\phi$ in a $\chi$ thermal bath can be computed using standard thermal field theory (see, e.g., \cite{Quiros:1999jp}). The result is given by:
\begin{align}
V_{\rm full}(\phi,T) &= - \frac{g_\chi}{2\pi^2}T^4 \int_0^\infty {\rm d}y\,y^2 \log\left[1+e^{-\sqrt{y^2+\mchieff^2/T^2}}\right]+\frac{1}{2}m_\phi^2\phi^2 \equiv \frac{g_\chi}{2\pi^2} m_\chi^4 {\cal V}_{\rm full}(\varphi,x)\;,\label{eq:Vfull} 
\end{align}
with the dimensionless potential defined as
\begin{align}
{\cal V}_{\rm full}(\varphi,x)  &= -\frac{1}{x^4} \int_0^\infty {\rm d}y\,y^2 \log\left[1+e^{-\sqrt{y^2+\gamma^2 x^2}}\right] +\kappa\frac{\varphi^2}{2}\;,\label{eq:Vfullnormalized}   
\end{align}
where $\gamma(\varphi) \equiv |1-\varphi^2/2|$.
Expanding Eq.~(\ref{eq:Vfull}) in the relativistic limit $\gamma x \ll1$ and keeping terms up to ${\cal O}(\varphi^2)$, one precisely reproduces the corresponding leading-order result in Eq.~(\ref{eq:FDrel}), namely the thermal correction to the quadratic term.
This confirms that the quadratic approximation is valid for our analysis in the crossover regime, where the decoupling occurs well before WIMP freeze-out and the condition $\gamma x\ll 1$ is always satisfied.
\vspace{0.2cm}

In the FOPT regime, decoupling occurs much later, but the ALP remains trapped in the symmetry-breaking vacuum, yielding $\gamma x \approx 1.33$. Consequently, the relativistic expansion still remains a good approximation. Fig.~\ref{fig:ThermalPotential} compares the full thermal potential (\ref{eq:Vfullnormalized}) with the approximate form (\ref{eq:potential}) used in the main text. The full potential likewise develops a symmetry-breaking minimum near $\varphi = \pm \sqrt{2}$ and, crucially, shows the characteristic degenerate vacua associated with a first-order transition at small $\kappa$. This confirms that the qualitative dynamics described in the main text, including the delayed symmetry restoration and the trapping of the ALP field, persist beyond the quadratic approximation. In particular, WIMP freeze-out can still be significantly delayed for sufficiently small $\kappa$.
Quantitatively, we find a slightly different scaling behavior of the coherent freeze-out temperature with respect to $\kappa$: $x_{\rm cfo}\sim \kappa^{-1/3}$ when using the full potential in Eq.~(\ref{eq:Vfullnormalized}) and $x_{\rm cfo}\sim \kappa^{-1/2}$ when using the approximate potential in Eq.~(\ref{eq:potential}). Consequently, for the same value of $\kappa \ll 1$, the full potential restores the symmetry at an earlier time compared to the approximate potential, leading to a milder enhancement of the relic abundance. In the numerical results shown in Figs.~\ref{fig:abundance} and \ref{fig:boostsigma} of the main text, we adopt the more accurate treatment based on the full potential.

\vspace{0.2cm}
In the following, we provide an analytical derivation of the power law $\xcfo \sim \kappa^{-1/3}$ with the full potential in Eq.~(\ref{eq:Vfullnormalized}). First, by changing the variable $y \to u\equiv\sqrt{y^2+\gamma^2 x^2}$, the integral in Eq.~(\ref{eq:Vfullnormalized}) becomes
\begin{align}
\int_0^\infty {\rm d}y\,y^2 \log\left[1+e^{-\sqrt{y^2+\gamma^2 x^2}}\right]
&=  \int_{\gamma x}^\infty {\rm d}u\,u\sqrt{u^2-\gamma^2 x^2}\,\log\left[1+e^{-u}\right]\nonumber\\
& = \sum_{n=1}^\infty \frac{(-1)^{n-1}}{n} \int_{\gamma x}^\infty {\rm d}u\,u\sqrt{u^2-\gamma^2 x^2}\, e^{-nu}\nonumber\\
& = \gamma^2x^2 \sum_{n=1}^\infty \frac{(-1)^{n-1}}{n^2} K_2(n\gamma x)\;.  
\end{align}
Then the full thermal potential can be written as
\begin{align}
{\cal V}_{\rm full}(\varphi,x)  &= -\frac{\gamma^2}{x^2} \sum_{n=1}^\infty \frac{(-1)^{n-1}}{n^2} K_2(n\gamma x) +\kappa\frac{\varphi^2}{2}\;.\label{eq:Vfullseries}  
\end{align}
As usual, we define $\xcfotilde$ as the time when the false vacuum $\varphi_*$ disappears,
\begin{align}
\frac{\partial{\cal V_{\rm full
}}(\varphi,x)}{\partial \varphi} \Bigg|_{\varphi=\varphi_*,\,x=\xcfotilde}=\frac{\partial^2{\cal V_{\rm full
}}(\varphi,x)}{\partial \varphi^2} \Bigg|_{\varphi=\varphi_*,\,x=\xcfotilde} = 0\;.  
\end{align}
Explicit calculation using Eq.~(\ref{eq:Vfullseries}) yields
\begin{align}
\frac{\partial{\cal V_{\rm full
}}(\varphi,x)}{\partial \varphi} =\varphi\left[\kappa - \sum_{n=1}^\infty\frac{(-1)^{n-1}}{n^3} \frac{c_n^2 K_1(c_n)}{x^3}
\right]&=0\;,\label{eq:D1Vfull}\\
\frac{\partial^2{\cal V_{\rm full
}}(\varphi,x)}{\partial \varphi^2} =\kappa-\sum_{n=1}^\infty \frac{(-1)^{n-1}}{n^3}\frac{c_n}{x^3} \left[2 c_n \left(n x -c_n\right)K_0(c_n) - \left(2 n x -3c_n\right)K_1(c_n) 
\right] &= 0\;,\label{eq:D2Vfull}  
\end{align}
where $c_n \equiv n\gamma x \equiv n c_1$. Note that the physical stationary point $\varphi_*$ can only approach $\sqrt{2}$ from below.
Substituting Eq.~(\ref{eq:D2Vfull}) into Eq.~(\ref{eq:D1Vfull}) to eliminate $\kappa$, one obtains
\begin{align}
\sum_{n=1}^\infty (-1)^{n-1}  \left[K_0(nc_1) -\frac{K_1(nc_1)}{nc_1}
\right]=0\;.\label{eq:Besselfunction-condition}  
\end{align}

Recall that the coherent freeze-out occurs as $c_1 = \gamma x \gtrsim {\cal O}(1)$, i.e., when the Bessel functions get Boltzmann suppressed, $K_\nu (nc_1) \sim e^{-nc_1}$. In this region, one can ignore all $n>1$ modes in Eq.~(\ref{eq:Besselfunction-condition}), obtaining
\begin{align}
 c_1 K_0(c_1) -K_1(c_1) =0 \quad \Rightarrow \quad c_1 \approx 1.33\;.    
\end{align}
Substituting it back to Eq.~(\ref{eq:D1Vfull}) yields
\begin{align}
\kappa = \frac{c_1^2}{\xcfotilde^3}\left[\sum_{n=1}^\infty (-1)^{n-1} \frac{K_1(n c_1)}{n}\right] \equiv S\frac{c_1^2}{\xcfotilde^3}\;,    
\end{align}
Using the numerical sum of the Bessel function:
\begin{align}
S \equiv  \sum_{n=1}^\infty (-1)^{n-1} \frac{K_1(n c_1)}{n} \approx 0.33\;,
\end{align}
we obtain
\begin{align}
\xcfotilde = S^{1/3}c_1^{2/3}\,\kappa^{-1/3} \approx 0.83\,\kappa^{-1/3}\;.    
\end{align}
This reproduces the correct scaling behavior used in Eq.~(\ref{eq:xcfotildekappa}) of the main text. It also agrees very well with the numerical solution (see Fig.~\ref{fig:xcfovskappa}).

\end{appendix}

\clearpage
\onecolumngrid
\begin{center}
    \vspace*{\fill}
    \includegraphics[width=0.8\textwidth]{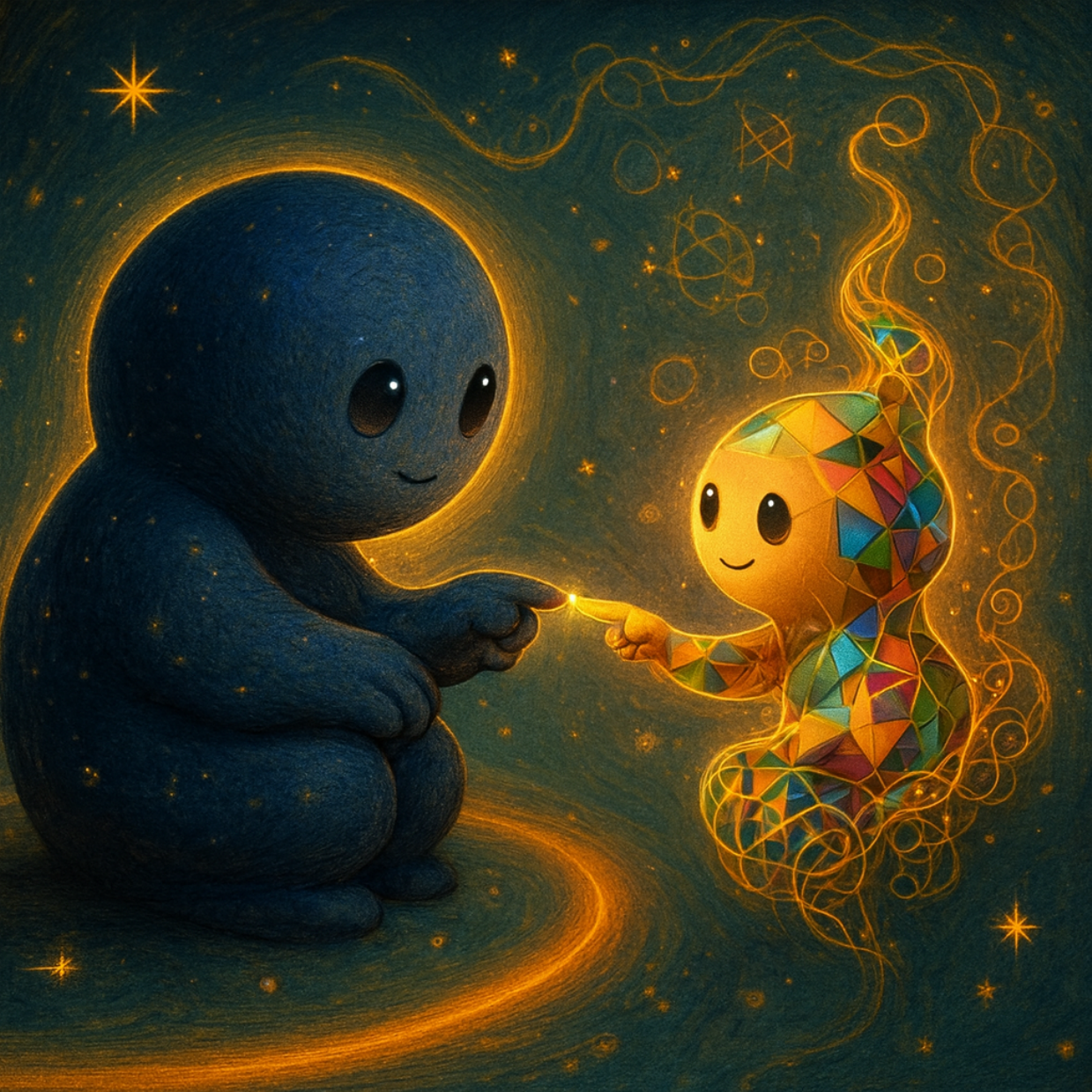}

    \bigskip

    {\Large\bfseries The story of ``WIMP meets ALP"}

    \vspace*{\fill}
\end{center}
\twocolumngrid
\end{document}